# Critical Ambient Pressure and Critical Cooling Rate in Optomechanics of Electromagnetically Levitated Nanoparticles

**AMIR M. JAZAYERI**

*Department of Electrical Engineering, Sharif University of Technology, Tehran 145888-9694, Iran*
*amirjazaayeri@gmail.com*

**Abstract:** The concept of critical ambient pressure is introduced in this paper. The particle escapes from its trap when the ambient pressure becomes comparable with or smaller than a critical value, even if the particle motion is cooled by one of the feedback cooling (or cavity cooling) schemes realized so far. The critical ambient pressure may be so small that it is not a limiting factor in ground-state cooling, but critical feedback cooling rates, which are also introduced in this paper, are limiting factors. The particle escapes from its trap if any of the feedback cooling rates (corresponding to the components of the particle motion) becomes comparable with or larger than its critical value. Critical feedback cooling rate is different from the well-known manifestation of the measurement noise. The critical feedback cooling rate corresponding to a certain component of the particle motion is usually smaller than the optimum feedback cooling rate at which the standard quantum limit happens unless that component is cooled by the Coulomb force (instead of the optical gradient force). Also, given that the measurement noise for the *z* component of the particle motion is smaller than the measurement noises for the other two components (assuming that the beam illuminating the particle for photodetection propagates parallel to the *z* axis), the feedback scheme in which the *z* component of the particle motion is cooled by the Coulomb force has the best performance. This conclusion is in agreement with the experimental results published after writing the first version of this paper. The dependence of the critical ambient pressure, the critical feedback cooling rates, and the minimum achievable mean phonon numbers on the parameters of the system is derived in this paper, and can be verified experimentally. The insights into and the subtle points about the EM force (including the gradient force, radiation pressure, and recoil force), the EM force fluctuations, and the measurement noise that are presented in this paper are all of theoretical and practical importance, and might be useful in many systems besides those examined in this paper.

## 1. Introduction

Cavity optomechanics of non-levitated objects examines the interaction between the EM fields of a resonator and the vibrations of its body [1-11]. The vibrations are usually a standing wave [1-4,7-11], and sometimes a travelling wave [5,6]. Some conceivable applications of cavity optomechanics are amplification of microwave signals [12], mechanical memory [13], synchronization of mechanical oscillators [14,15], induced transparency [16], and optical frequency conversion [17]. However, such applications are subject to slow response of mechanical oscillators. Still, cavity optomechanics offers unique opportunities in the quantum regime [18-20], which necessitate cooling the mechanical vibrations to near their ground state [3,4]. Cavity cooling of mechanical oscillations parallels Doppler cooling in atomic physics; the former is thanks to the frequency-selective EM response of resonators, and the latter is thanks to the frequency-selective EM response of atoms. The other way of cooling the mechanical vibrations is to use feedback [21]; the idea is to continuously measure the velocity of the vibrating wall, and exert an optical force proportional to and in the opposite direction of its velocity on it. The minimum mean phonon number (in steady state) achieved by cavity cooling or feedback cooling is a decreasing function of the intrinsic damping rate of the mechanical vibrations, because, as a manifestation of the fluctuation-dissipation theorem, the spectral density of the exerted thermal force on the vibrating part is proportional to the intrinsic damping rate of its vibrations. Design and fabrication of structures with mechanical modes of extremely small intrinsic damping rates is in fact a major challenge in optomechanics of non-levitated objects.

To avoid the challenge of engineering the mechanical characteristics of the system, one idea is to use an electromagnetically levitated particle as the mechanical part of the system, because the intrinsic damping rate of its oscillations around the trapping point can be made arbitrarily small simply by reducing the ambient pressure [22]. In cavity cooling systems, the cavity mode supposed to cool the particle motion may be excited directly by a laser [23-25], or excited by the photons scattered by the particle [26-28]. In feedback cooling systems, the force cooling the particle motion may be an optical force [29-33], or a combination of an optical force and a Coulomb force [34-36] (in the latter case, the particle has to have a net charge).

In Section 3, it will be shown that the particle escapes from its trap if the intrinsic damping rate (or, equivalently, the ambient pressure) becomes comparable with or smaller than a critical value. The critical value comes from the inherent uncertainty in the emission of photons by the laser trapping the particle, and the resulting fluctuations in the trapping force, which is linear in the position of the particle (viz., its corresponding Hamiltonian is quadratic in the position of the particle). It is worth noting that the usual Hamiltonian in optomechanics is linear in the position of the mechanical part of the system [1-4,8-10,23-28], and this is why the Heisenberg equations (which are always non-linear) is usually linearized in optomechanics, whether in the weak [1,2,4,23-28] or strong optomechanical interaction regime [3,8-10]. However, the Hamiltonian corresponding to the trapping force in our problem is quadratic in the position operator, and the resulting Heisenberg equations cannot be linearized. To derive the critical value of the intrinsic damping rate, we use a quantum noise approach like the one in [11]; however, unlike [11], a simple closed-form solution to the steady-state mean phonon number will be found in our problem. After writing this paper, we realized that a similar (and not exactly the same) analysis of the trapping force fluctuations had previously been presented in the context of atom traps [37,38], which has seldom been cited by papers on optomechanics. What is called 'heating rate' in [37,38] is similar to the rate $8\Gamma_{g,i}$ derived in Section 3, where we will argue that the term 'heating rate' might be misleading.

The dependence of the critical intrinsic damping rate (and its corresponding critical ambient pressure) on the parameters of the system will be derived in this paper. Also, it will be argued that the cavity cooling and feedback cooling schemes realized so far cannot counteract the particle's escape at ambient pressures comparable with or smaller than the critical value. It should be noted that the critical ambient pressure may be so small that it is not a limiting factor in ground-state cooling, but critical feedback cooling rates (which have the same origin as the critical ambient pressure, and will be derived in Section 4) are limiting factors. The concept of critical feedback cooling rate has not so far been introduced in the literature.

In Section 4, the parametric feedback cooling scheme [31-33] (which cools the particle motion by the optical gradient force) and the hybrid feedback cooling scheme [34-36] (which cools two components of the particle motion by the optical gradient force, and the other component by the Coulomb force) will be analyzed. It will be discussed that the measurement noise coming from the uncertainty in the emission of photons by the laser illuminating the particle for photodetection (which is usually the same as the laser trapping the particle) have two manifestations. One manifestation is well-known, and is relevant to feedback cooling systems of levitated particles [34] and non-levitated objects [21] both. The other manifestation, which is relevant only to feedback cooling systems of levitated particles, and has not so far been investigated in the literature, is the existence of critical feedback cooling rates. The particle escapes from its trap if any of the feedback cooling rates (corresponding to the components of the particle motion) becomes comparable with or larger than its critical value. The critical feedback cooling rates will be derived by using a novel self-consistent method.

In section 4, it will also be derived how the minimum achievable mean phonon number (for each component of the particle motion) depends on the parameters of the system, especially on the mean laser power used to illuminate the particle for photodetection (which is usually the same as the mean laser power used to trap the particle), the radius of the particle, and the numerical aperture of the lens employed to generate the beam illuminating the particle for photodetection (which is usually the same as the beam trapping the particle). In Section 5, by giving some numerical examples, it will be demonstrated whether the parametric feedback cooling scheme and the hybrid feedback cooling scheme are able to cool the components of the particle motion to near their ground states. Our conclusions are in agreement with the experimental results reported in the literature.

It should be noted that there is a realization of feedback cooling of levitated particles which was reported in [30], and will not be analyzed in Section 4. This feedback cooling scheme can be analyzed in a way similar to the analyses presented in Section 4, but it should be noted that the expressions of the critical feedback cooling rates for this scheme are completely different from the expressions given in Section 4. The analysis of this feedback cooling scheme was removed from the current version of the paper, because this scheme is affected by practical issues, and is no longer used; in this scheme, the components of the cooling force are generated by three lasers distinct from the laser trapping the particle (and illuminating it for photodetection), and therefore, the misalignment of the axes of the cooling beams with respect to the coordinate axes (defined by the trapping beam) is a serious issue.

Finally, it is worth noting that the points which will be made in the main text as well as the appendices about the EM force (including the gradient force, radiation pressure, and recoil force), the EM force fluctuations, and the measurement noise might be useful in any system employing levitated particles (e.g. the system reported in [39]). Also, we note that not only a systematic study of the EM force fluctuations has not so far been presented in the literature, but also there even exist misconceptions about the classical EM force (e.g. see [40] which tries to clear up some of those misconceptions).

## 2. Effects of gas molecules

Let us consider a small dielectric particle of mass $M$ levitated by the optical gradient force around the focal point of a lens whose axis is defined as the $z$ axis. I denote the position (or position operator) of the particle center with respect to the focal point of the lens by $\vec{r} = (x_1, x_2, x_3)$. The gas molecules surrounding the particle exert a damping force $-M\Gamma\dot{\vec{r}}$ on it, where the coefficient $\Gamma$, which will hereafter be called ‘the intrinsic damping rate’, is proportional to the ambient pressure $P_{am}$ (see [41] and Appendix C). As a manifestation of the fluctuation-dissipation theorem, the gas molecules also exert a random force $\vec{f}_{th}$ on the particle, where the spectral density of $f_{th,i}$ is proportional to $\Gamma$ (see [42] for the definition of spectral density). Ignoring the laser power fluctuations for the moment, the variance of $x_i$ (see [43] for the definition of variance) can be written as $(2\overline{m}_{th,i}+1)\hbar/(2M\Omega_i)$, where the mechanical oscillation frequency $\Omega_i$ is determined by the optical gradient force, and the mean phonon number $\overline{m}_{th,i}$, which reads $[\exp(\frac{\hbar\Omega_i}{k_B T})-1]^{-1}$, can be approximated by $k_B T/(\hbar\Omega_i)$ for $k_B T \gg \hbar\Omega_i$. The temperature $T$ in the expression of $\overline{m}_{th,i}$ has a value between the ambient temperature and the surface temperature of the particle (see [44,45] and Appendix C). It is noteworthy that the photophoretic force, which is a result of temperature gradient over the particle surface [46,47], is negligible in our problem.

## 3. EM force and its fluctuations

The exerted EM force on the particle can be written as the sum of three terms called gradient force, radiation pressure, and recoil force (see Appendices A and B). The gradient force comes from the dependence of the EM energy on the position of the particle. In other words, the gradient force is exerted on the particle by the EM modes *included* in the Hamiltonian. In contrast, interaction between the *system* (consisting of the particle and the EM modes included in the Hamiltonian) and the EM modes *excluded* from the Hamiltonian leads to radiation pressure and the recoil force. Radiation pressure comes from the initial momentum of the photons scattered or absorbed by the particle, while the recoil force comes from their final momentum. Since our dielectric particle has a low loss, we can ignore the contribution of the initial momentum of the photons absorbed by the particle in radiation pressure.

It is noteworthy that in cavity optomechanics of levitated particles [23-28], the force trapping the particle as well as the force cooling the particle motion is in fact a gradient force. Also, in cavity optomechanics of non-levitated objects [1,2], where the EM fields of a resonator interact with the vibrations of the body of the resonator, the force cooling the vibrations is usually called 'radiation pressure', but it is in fact a conservative force (viz., it is exerted by the EM modes *included* in the Hamiltonian), and is therefore similar to the gradient force in this paper.

In this section, I examine quantum fluctuations in the exerted trapping force on the particle, and show how they lead to a critical value for the ambient pressure. It should be noted that unlike the usual Hamiltonian in optomechanics [1-4,8-10,23-28], which is linear in the position of the movable part of the system, the Hamiltonian corresponding to the trapping force is quadratic in the position of the particle. Also, it should be noted that the trapping EM field in our problem is non-resonant, but a similar critical ambient pressure can be derived for the schemes whose trapping EM field is resonant. The to-be-derived critical ambient pressure is usually so small that it is not a limiting factor in ground-state cooling, but we shall see in the next section that fluctuations in the gradient force also lead to a critical feedback cooling rate which is a limiting factor. In the following, I also take into account quantum fluctuations in the exerted radiation pressure and recoil force on the particle.

Our *system* consists of the particle and a Gaussian beam. The Gaussian beam around the focal point of the lens can be considered as a valid EM mode satisfying Maxwell's equations [48,49]. The exerted gradient force $(g_1, g_2, g_3)$ on the particle acts as a spring force $-(K_1x_1, K_2x_2, K_3x_3)$ around the focal point of the lens, and allows us to define mechanical modes with the oscillation frequencies $\Omega_i = \sqrt{K_i / M}$ . Let us write $K_i$ as $A_i P_L$, where $P_L$ denotes the power carried by the Gaussian beam, and $A_i$ are coefficients given in Appendix A. We can infer from the expression $g_i = -A_i P_L x_i$, which is derived in the framework of classical electrodynamics, that the Hamiltonian corresponding to the gradient force component $g_i$ reads $A_i P_L x_i^2 / 2$ , in which $P_L$ and $x_i$ are interpreted as operators. In Appendix B, I will discuss that this inference is not exactly accurate, but can be used in our problem (and similar problems), where we can ignore the effect of the particle motion on the evolution of the non-resonant EM mode included in the Hamiltonian. However, that inference would be exactly accurate if the EM mode was resonant, and the expression of the gradient force was written in terms of the energy of the EM mode.

Assuming that the laser power fluctuations are solely due to inherent uncertainty in the emission of photons by the laser, the spectral density of $P_L$ (more accurately, the spectral density of $P_L$ in the absence of the particle) is the constant function $S_{P_L}(\omega) = \hbar \omega_0 \overline{P}_L$, where $\overline{P}_L$ denotes the expectation value of $P_L$ (see [42] for the definition of spectral density). It is noteworthy that if the EM mode was resonant, its energy would appear in the gradient force and the corresponding Hamiltonian; however, the spectral density of the EM mode energy (in the absence of the particle) would not be a constant function of $\omega$ .

The exerted radiation pressure on the particle around the focal point of the lens can be written as $\hat{x}_3 \rho_3$ (viz., $\rho_1, \rho_2 \equiv 0$), where $\hat{x}_3$ denotes the unit vector parallel to the *z* axis. Also, $\rho_3$ is almost insensitive to the position of the particle (see Appendices A and B). The expectation value of $\rho_3$ can be derived in the framework of classical electrodynamics, and written as $B\overline{P}_L$ , where $B$ is a coefficient given in Appendix A, and $\overline{P}_L$ was defined above. Interestingly, the quantum treatment of radiation pressure also yields exactly the same result for the expectation value of $\rho_3$ (see Appendix B). However, we can never infer from $\overline{\rho}_3 = B\overline{P}_L$ that the Hamiltonian corresponding to radiation pressure reads $-BP_L x_3$, because $-BP_L x_3$ does not capture the role of the EM modes *excluded* from the system. As a result, the spectral density of $\rho_3$ is not equal to $B^2 S_{P_L}(\omega)$. Rather, the spectral density of $\rho_3$ can be written as $B'^2 S_{P_L}(\omega)$ , where $B'$, which is given in Appendix B, is not equal to $B$. It should be noted that if the laser power fluctuations were mainly due to fluctuations in the electric current applied to the laser [viz., if $S_{P_L}(\omega)$ , apart from a coefficient, was equal to the spectral density of the electric current applied to the laser], the spectral density of $\rho_3$ would be equal to $B^2 S_{P_L}(\omega)$ . We will return to this point when we discuss feedback cooling in the next section.

Unlike radiation pressure, the recoil force has a zero expectation value. In other words, the presence of the recoil force is not deducible from classical electrodynamics. The recoil force has three components $\hat{x}_i \sigma_i$ (for *i*=1,2,3), and is almost insensitive to the position of the particle. The spectral density of $\sigma_i$ reads $C_i^2 \hbar \omega_0 \overline{P}_L$ , where $C_i$ is given in Appendix B. As is discussed therein, even if there were fluctuations in the electric current applied to the laser, the spectral density of $\sigma_i$ could still be approximated by $C_i^2\hbar\omega_0\overline{P}_L$. This point will be used when we discuss feedback cooling in the next section.
It is noteworthy that as a manifestation of the fluctuation-dissipation theorem, the radiation pressure and recoil force fluctuations are accompanied by a damping force. However, this damping force, which can also be explained by the Doppler effect [50,51], is negligible (at least in our problem).

To examine the time evolution of the mechanical state of the system, let us use a quantum noise approach similar to the one in cavity optomechanics of non-levitated objects in the weak optomechanical coupling regime [1,2,11]. We assume that the spectral density of $x_i$ can be approximated by a Lorentzian function whose central frequency is $\Omega_i$ (viz., with peaks at $\omega = \pm\Omega_i$). Also, let us assume that the particle motion does not significantly change the spectral density of the power carried by the EM mode (viz., the Gaussian beam) interacting with the particle. Fermi's golden rule can then be used to find the transition rates between the mechanical Fock states. It is noteworthy that the use of Fermi's golden rule is in fact equivalent to expanding the time evolution of the reduced density matrix (viz., the density matrix after tracing over the Gaussian beam and the thermal bath) to second order in $P_L - \overline{P}_L$ and $f_{th,i}$. The quantum noise approach leads to the following rate equation:

$$
\begin{aligned}
\dot{P}_{i,m} &= \Gamma_{g,\uparrow,i}m(m-1)P_{i,m-2} + (\overline{m}_{th,i}\Gamma + \Gamma_{r,\uparrow,i})mP_{i,m-1} \\
&-(\overline{m}_{th,i}\Gamma + \Gamma_{r,\uparrow,i})(m+1)P_{i,m} - [\overline{m}_{th,i}\Gamma + \Gamma + \Gamma_{r,\downarrow,i}]mP_{i,m} \\
&-\Gamma_{g,\uparrow,i}(m+2)(m+1)P_{i,m} - \Gamma_{g,\downarrow,i}(m-1)mP_{i,m} \\
&+[\overline{m}_{th,i}\Gamma + \Gamma + \Gamma_{r,\downarrow,i}](m+1)P_{i,m+1} + \Gamma_{g,\downarrow,i}(m+1)(m+2)P_{i,m+2},
\end{aligned}
\tag{1}
$$

where $P_{i,m}(t)$ denotes the probability that the mean phonon number $m_i$ (for the *i*th component of the particle motion) at *t* equals *m*. The *i*th component of the thermal force fluctuations (viz., $f_{th,i}$) leads to a transition rate $\overline{n}_{th,i}\Gamma(m+1)$ from the mechanical Fock state $\left|m\right\rangle_i$ to $\left|m+1\right\rangle_i$, and a transition rate $\overline{n}_{th,i}\Gamma m$ from $\left|m\right\rangle_i$ to $\left|m-1\right\rangle_i$. The *i*th component of the damping force (viz., $-\hat{x}_i M\Gamma\dot{x}_i$) leads to a transition rate $\Gamma m$ from $\left|m\right\rangle_i$ to $\left|m-1\right\rangle_i$ only. The *i*th components of the radiation pressure fluctuations (viz., $\rho_i - \overline{\rho}_i$) and the recoil force fluctuations (viz., $\sigma_i$) lead to a transition rate $\Gamma_{r,\uparrow,i}(m+1)$ from $\left|m\right\rangle_i$ to $\left|m+1\right\rangle_i$, and a transition rate $\Gamma_{r,\downarrow,i}m$ from $\left|m\right\rangle_i$ to $\left|m-1\right\rangle_i$, where $\Gamma_{r,\uparrow,i}$ and $\Gamma_{r,\downarrow,i}$ read $[S_{\rho_i}(-\Omega_i) + S_{\sigma_i}(-\Omega_i)]/(2\hbar M\Omega_i)$ and $[S_{\rho_i}(\Omega_i) + S_{\sigma_i}(\Omega_i)]/(2\hbar M\Omega_i)$, respectively, in terms of the spectral densities of $\rho_i$ and $\sigma_i$ calculated at $-\Omega_i$ and $\Omega_i$. Since $\Gamma_{r,\uparrow,i}$ and $\Gamma_{r,\downarrow,i}$ are equal, they will hereafter be denoted by $\Gamma_{r,i}$. The *i*th component of the trapping force fluctuations [viz., $-A_i(P_L - \overline{P}_L)x_i$] leads to a transition rate $\Gamma_{g,\uparrow,i}(m+1)(m+2)$ from $\left|m\right\rangle_i$ to $\left|m+2\right\rangle_i$, and a transition rate $\Gamma_{g,\downarrow,i}m(m-1)$ from $|m\rangle_i$ to $|m-2\rangle_i$, where $\Gamma_{g,\uparrow,i}$ and $\Gamma_{g,\downarrow,i}$ read $A_i^2 S_{P_L}(-2\Omega_i)/(4M\Omega_i)^2$ and $A_i^2 S_{P_L}(2\Omega_i)/(4M\Omega_i)^2$, respectively, in terms of the spectral density of $P_L$ calculated at $-2\Omega_i$ and $2\Omega_i$, respectively. Since $\Gamma_{g,\uparrow,i}$ and $\Gamma_{g,\downarrow,i}$ are equal, they will hereafter be denoted by $\Gamma_{g,i}$. It is noteworthy that $\Gamma_{g,\uparrow,i}$ and $\Gamma_{g,\downarrow,i}$ would not be equal if the EM mode was resonant (viz., if it was supported by a resonator).

The mean phonon number is by definition equal to $\overline{m}_i(t) = \sum_m m P_{i,m}(t)$. We are interested in steady state, viz., at large enough *t*, where $\dot{P}_{i,m}(t)$ all vanish, and the spectral density of $x_i$ is definable. By using Eq. (1), we find that the steady-state mean phonon number, which is usually shortened to 'mean phonon number', has the following closed-form solution

$$\overline{m}_i = (\overline{m}_{th,i}\Gamma + \Gamma_{r,i} + 4\Gamma_{g,i}) / (\Gamma - 8\Gamma_{g,i}) . \qquad (2)$$

The three terms in the numerator of $\overline{m}_i$ are due to the thermal force fluctuations, the recoil force fluctuations (together with the radiation pressure fluctuations), and the trapping force fluctuations, respectively. The first term in the denominator of $\overline{m}_i$ is thanks to the damping force $-\hat{x}_i M\Gamma\dot{x}_i$, which counteracts the effect of the force fluctuations. Note that, in the absence of the EM force fluctuations, $\overline{m}_i$ reduces to $\overline{m}_{th,i}$. The second term in the denominator of $\overline{m}_i$ is due to the *renormalization* of the susceptibility of $x_i$ by the trapping force fluctuations. The expression of $\overline{m}_i$ indicates that the intrinsic damping rate ($\Gamma$) must be kept well above the critical value $\Gamma_{cr} = \max_i(8\Gamma_{g,i})$ – otherwise, the renormalization effect becomes so strong that the particle escapes from the trap. Therefore, it is impossible to trap the particle in vacuum even by an ideal laser whose field fluctuations are solely due to inherent uncertainty in the emission of photons. After writing this paper, we realized that a similar (and not exactly the same) analysis of the trapping force fluctuations had previously been presented in the context of atom traps [37,38], which has seldom been cited in papers on optomechanics. The mean phonon number has not been derived in [37,38], but what is called 'heating rate' in [37,38] is similar to the rate $8\Gamma_{g,i}$ derived above. The term 'heating rate' might be misleading, because one might mistakenly think that the 'heating rate' in [37,38] leads to terms such as those that appeared in the 'numerator' of the mean phonon number in Eq. (2), whereas the 'heating rate' in [37,38] in fact acts as a negative damping rate.

Unlike the usual renormalization effect in cavity optomechanics [1,2,4], which is mainly due to an asymmetry in the spectral density of the energy of a resonant EM mode (and the resultant asymmetry in the spectral density of an EM force which has *no explicit dependency* on the position operator), the renormalization effect discussed above is due to the explicit dependency of the trapping force on $x_i$. Also, it should be noted that the spectral density of $x_i$ can no longer be approximated by a Lorentzian function with a well-defined central frequency when $\Gamma$ becomes comparable to (or smaller than) $\Gamma_{cr}$, because the trapping force fluctuations directly affect the stiffness of the spring defining the mechanical mode of the oscillation frequency $\Omega_i$. In other words, when $\Gamma$ becomes comparable to (or smaller than) $\Gamma_{cr}$, at least one of the mechanical modes disappears. This destruction of the mechanical modes by the trapping force fluctuations is to some extent similar to the destruction of the Higgs mode [52] in the magnetically ordered phase of the quantum rotor model in low dimensions.

In deriving Eq. (2), the intrinsic damping force $-M\Gamma\dot{\vec{r}}$ has been assumed to be the only force cooling the particle motion. If an *ideal* cooling force whose *i*th component reads $F_i = -M\Gamma_{c,i}\dot{x}_i$ was also exerted on the particle, $\Gamma$ in the denominator of $\overline{m}_i$ would be replaced by $\Gamma + \Gamma_{c,i}$. In such a case, if $\Gamma_{c,i} \gg \Gamma_{cr}$, we would not need to keep $\Gamma$ well above $\Gamma_{cr}$. However, the cooling schemes which have been realized so far do not produce the ideal cooling force, and do not allow us to remove the constraint that $\Gamma$ must be kept well above $\Gamma_{cr}$, whether they are cavity-based or feedback-based. Cavity cooling in general [1-4,8-10] and cavity cooling of levitated particles in particular [23-28] can only cool well-defined mechanical modes efficiently – in other words, the spectral density of $x_i$ has to be a function with a well-defined and specific central frequency, which is not the case when $\Gamma$ becomes comparable with (or smaller than) $\Gamma_{cr}$. As to the feedback cooling schemes which have been realized so far [31-36], we will discuss in the next section that they require that the spectral densities of $x_j^2$ and $x_i^2$ (for $i \neq j$) do not overlap with each other, which is not the case when $\Gamma$ becomes comparable with (or smaller than) $\Gamma_{cr}$. It is worth emphasizing that 'spectral density', which is a steady state quantity, is not definable in the first place when $\Gamma$ becomes comparable with (or smaller than) $\Gamma_{cr}$, whether the particle motion is cooled by one the cooling schemes which have been realized so far or not.

Interestingly, $\Gamma_{cr}$ is insensitive to $\overline{P}_L$ and also to the radius (*R*) of the particle, because $\Gamma_{g,i}$ (for all *i*) is insensitive to $\overline{P}_L$ and *R*. Also, $\Gamma_{cr}$ is an increasing function of the numerical aperture (*NA*) of the lens employed to generate the trapping beam. The critical ambient pressure $P_{am,cr}$ corresponding to $\Gamma_{cr}$ is (almost) insensitive to $\overline{P}_L$, but is proportional to *R*, and is an increasing function of *NA* (see Appendix C). As is usually the case in the feedback systems of levitated particles [31,32,34-36], let us assume that the beam trapping the particle illuminates it for photodetection as well – otherwise [33], the contribution of the illuminating beam in the rates $\Gamma_{r,i}$ and $\Gamma_{g,i}$ (and the critical value $\Gamma_{cr}$) can be simply taken into account.

## 4. Feedback cooling

The idea of feedback cooling is to measure $\dot{\vec{r}}$, and generate a cooling force whose *i*th component is $F_i = -M\Gamma_{fb,i}\dot{x}_i$ [21]. This section examines the parametric feedback cooling scheme [31-33] as well as the hybrid feedback cooling scheme [34-36]. In the parametric feedback cooling scheme, the electric currents carrying the necessary information about the particle motion are applied to the laser trapping the particle. In the hybrid feedback cooling scheme, the electric current carrying the necessary information about one component of the particle motion is applied to a capacitor, while the electric currents carrying the necessary information about the other two components are applied to the laser trapping the particle.

Two major issues are to be discussed in this section. The first one is what I name the issue of 'unwanted force components'. We will see that each component of the cooling force is accompanied by unwanted force components which disable feedback cooling if there are overlaps between the spectral densities of the components of the particle motion. This is why neither parametric feedback cooling nor hybrid feedback cooling can counteract the destruction of the mechanical modes by the trapping force fluctuations discussed in the previous section. In other words, the intrinsic damping rate $\Gamma$ must be kept well above the critical value $\Gamma_{cr}$ (derived in the previous section) even in the presence of feedback cooling. It is worth emphasizing that the issue of unwanted force components is not relevant to feedback systems of non-levitated objects (e.g. the feedback system reported in [21]).

The second issue to be discussed in this section is the measurement noise. We saw in the previous section that inherent uncertainty in the emission of photons by the laser trapping the particle as well as the laser illuminating the particle for photodetection (which are usually the same) leads to the rates $\Gamma_{r,i}$ and $\Gamma_{g,i}$ (and the critical value $\Gamma_{cr}$ ). However, the uncertainty in the emission of photons by the laser illuminating the particle also leads to the measurement noise: the photodetector intended to measure $x_i$ (for $i$=1,2,3) generates a photocurrent $I_i$ whose spectral density, apart from an unimportant coefficient, can be written as $S_{x_i}(\omega)+S_{n_i}(\omega)$ . We can interpret $I_i$ (apart from the unimportant coefficient) as an incoherent sum of $x_i$ and $n_i$, and write it as $x_i \oplus n_i$. In Appendix D, subtle assumptions and approximations involved in deriving $S_{n_i}(\omega)$ are discussed. Also, $S_{n_i}(\omega)$, which is a constant function of $\omega$ , is derived in terms of the parameters I name 'effective distance' and 'effective area'. It is noteworthy that the variance of $n_i$ is much larger than the variance of $x_i$, but the variance of $n_i$ as seen by the particle (viz., considering the small linewidth of the susceptibility of $x_i$ ) is usually much smaller than the variance of $x_i$ – hence the name 'noise' for $n_i$.

The measurement noise is not peculiar to feedback systems of levitated particles. Feedback systems of non-levitated objects (e.g. the feedback system reported in [21]) are also afflicted by the measurement noise. However, the measurement noise in feedback systems of levitated particles has two manifestations. The first manifestation, which is also common to feedback systems of non-levitated objects, is an increase in the numerator of Eq. (2). The second manifestation, which is the one that is absent in feedback systems of non-levitated objects, is a decrease in the denominator of Eq. (2). More precisely, the mean phonon number $\overline{m}_i$ (for $i$=1,2,3) in the presence of feedback cooling can be written as

$$\overline{m}_i=(\overline{m}_{th,i}\Gamma+\Gamma_{r,i}+4\Gamma_{g,i}+\tilde{\Gamma}_{r,i}+4\tilde{\Gamma}_{g,i})/(\Gamma+\Gamma_{fb,i}-8\Gamma_{g,i}-8\tilde{\Gamma}_{g,i}), \tag{3}$$

where $\Gamma_{g,i}$ and $\Gamma_{r,i}$ were derived in the previous section, and $\tilde{\Gamma}_{g,i}$ and $\tilde{\Gamma}_{r,i}$ will be derived in this section for different realizations of the feedback system.

In view of Eq. (3), one might think that $\Gamma + \Gamma_{fb,i}$ is the quantity which must be kept well above $8\Gamma_{g,i} + 8\tilde{\Gamma}_{g,i}$. However, due to the first issue (viz., the issue of 'unwanted force components'), neither parametric feedback cooling nor hybrid feedback cooling can counteract the destruction of the mechanical modes. In other words, the intrinsic damping rate $\Gamma$ (not $\Gamma + \Gamma_{fb,i}$) is the quantity which must kept well above $8\Gamma_{g,i} + 8\tilde{\Gamma}_{g,i}$. The condition that $\Gamma$ must be kept well above $8\tilde{\Gamma}_{g,i}$ (for all *i*) means that the feedback cooling rates must be kept well below critical values which will be derived in this section. Assuming that $\Gamma$ is much larger than $8\Gamma_{g,i} + 8\tilde{\Gamma}_{g,i}$, Eq. (3) can be simplified to

$$\overline{m}_i = (\overline{m}_{th,i}\Gamma + \Gamma_{r,i} + \tilde{\Gamma}_{r,i}) / (\Gamma + \Gamma_{fb,i}) . \qquad (4)$$

### *4.1 Parametric feedback cooling*

We now examine the parametric feedback cooling scheme [31-33], in which the cooling force is generated by the laser trapping the particle. The electric current $I_c$ which carries the necessary information about the velocity of the particle, and is to be applied to the laser, can be written as $\overline{I}_c + \delta I_c$, where the electric current fluctuations $\delta I_c$ contain the useful information and the measurement noise both. Since $I_c$ is much smaller than the current $\overline{I}_B$ that provides the laser power used to trap (and illuminate) the particle, we can ignore the uncertainty in the emission of photons by the laser when we consider the laser power fluctuations coming from $I_c$. In other words, unlike the laser power fluctuations coming from $\overline{I}_B$, which was discussed in Section 3 and is due to the uncertainty in the emission of photons by the laser, we can assume that the laser power fluctuations coming from $I_c$ is linearly proportional to $\delta I_c$. In short, we can write the laser power operator as $P_L = \overline{P}_L + \delta P_l + \delta P_c$, where the spectral density of the power fluctuations $\delta P_l$ reads $\hbar\omega_0 \overline{P}_L$, and the power fluctuations $\delta P_c$ is linearly proportional to the electric current fluctuations $\delta I_c$.

The *i*th component of the gradient force operator can be readily written as $-A_i(\overline{P}_L + \delta P_l + \delta P_c)x_i$ in terms of the operators $\delta P_l$, $\delta P_c$, and $x_i$, where the coefficient $A_i$ was defined in Section 3, and is given in Appendix A. We saw in Section 3 that the radiation pressure operator and the recoil force operator (which are insensitive to the position of the particle) cannot be written solely in terms of the EM modes included in the Hamiltonian (viz., the Gaussian beam in our system). We saw that the spectral density of the radiation pressure fluctuations coming from $\delta P_l$ reads $B'^2\hbar\omega_0\overline{P}_L$ (viz., is proportional to the spectral density of $\delta P_l$ with the proportionality constant $B'^2$), and the spectral density of the *i*th component of the recoil force fluctuations coming from $\delta P_l$ reads $C_i^2\hbar\omega_0\overline{P}_L$ (viz., is proportional to the spectral density of $\delta P_l$ with the proportionality constant $C_i^2$), where the coefficients $B'$ and $C_i$ are given in Appendix B. However, as is discussed in Appendix B, the spectral density of the radiation pressure fluctuations coming from $\delta P_c$ is proportional to the spectral density of $\delta P_c$ with the proportionality constant $B^2$, where the coefficients $B$ and $B'$ are unequal. Also, as is discussed in Appendix B, we can ignore the recoil force fluctuations coming from $\delta P_c$.

The photodetector intended to measure $x_j$ (for *j*=1,2,3) generates a photocurrent $I_j$ which (apart from an unimportant coefficient) can be interpreted as an incoherent sum $x_j \oplus n_j$, where the spectral density $S_{n_i}(\omega)$ of the unwanted term $n_j$ is a constant function of $\omega$, and is derived in Appendix D.

To generate the *j*th component of the cooling force (viz., $-M\Gamma_{fb,j}\dot{x}_j$), one might think that we have to apply the current $\dot{I}_j$ (viz., the derivative of $I_j$) to the laser. However, since the exerted gradient force on the particle is linear in the position of the particle, the resulting force component would be of the form $-\eta_j x_j \dot{x}_j$, which does not cool the particle motion because its sign depends on $x_j$. Therefore, we have to multiply $\dot{I}_j$ by $I_j$ before applying it to the laser. The resulting force component is now of the form $-\eta_j x_j^2 \dot{x}_j$, which is a cooling force, but not of the desired form $-M\Gamma_{fb,j}\dot{x}_j$. Let us now introduce an approximation. Let us write $\eta_j$ as $M\Gamma_{fb,j} / \overline{x_j^2}$, and approximate $-\eta_j x_j^2 \dot{x}_j$ by $-M\Gamma_{fb,j}\dot{x}_j$, where, in general, $\overline{x_j^2}$ (viz., the expectation value of $x_j^2$) must be calculated self-consistently.

The cooling force component $-\eta_j x_j^2 \dot{x}_j$ (for each *j*) is accompanied by a gradient force $\hat{x}_i \tilde{g}_{ji}$ (for $i \neq j$), another gradient force $\hat{x}_i \tilde{g}_{n,ji}$ (for all *i*), and a radiation pressure $\hat{x}_3(\tilde{\rho}_{j3} + \tilde{\rho}_{n,j3})$, where $\tilde{g}_{ji}$, $\tilde{g}_{n,ji}$, $\tilde{\rho}_{j3}$, and $\tilde{\rho}_{n,j3}$ read $-(A_i / A_j)(\eta_j \dot{x}_j x_j) x_i$, $-(A_i / A_j)\eta_j (\dot{x}_j n_j \oplus x_j \dot{n}_j) x_i$, $(B / A_j)\eta_j \dot{x}_j x_j$, and $(B / A_j)\eta_j (\dot{x}_j n_j \oplus x_j \dot{n}_j)$, respectively, and the coefficients $A_i$ and $B$ were defined in Section 3, and are given in Appendix A. The forces which are proportional to $\dot{n}_j n_j$ have been ignored. For simplicity, let us define $a_{ji} \triangleq A_i / A_j$ and $b_j \triangleq B / A_j$.

The forces $\hat{x}_i \tilde{g}_{ji}$ (for $i \neq j$) and $\hat{x}_3 \tilde{\rho}_{j3}$, which are in fact what I named 'unwanted force components' at the beginning of this section, can potentially disable feedback cooling. Given that the unwanted force $\hat{x}_i \tilde{g}_{ji}$ (for $i \neq j$) reads $-\hat{x}_i a_{ji} \eta_j \dot{x}_j x_j x_i$, it is crucial that the spectral density of $\dot{x}_j x_j x_i$ (for $i \neq j$) does not overlap with the susceptibility of $x_i$. In other words, the spectral densities of $x_j^2$ and $x_i^2$ (for $i \neq j$) must not overlap with each other. Also, given that the unwanted force $\hat{x}_3\tilde{\rho}_{j3}$ reads $\hat{x}_3 b_j \eta_j \dot{x}_j x_j$, it is crucial that the spectral densities of $x_j^2$ (for all *j*) and $x_3$ do not overlap with each other. The unwanted force components are what make parametric feedback cooling unable to counteract the destruction of the mechanical modes (discussed in Section 3). Therefore, the intrinsic damping rate $\Gamma$ must be kept well above the critical value $\Gamma_{cr}$ (derived in Section 3) even in the presence of parametric feedback cooling. In short, we can say that the unwanted force $\hat{i}\tilde{g}_{ji}$ (for $i \neq j$) does not affect parametric feedback cooling if the conditions $\Gamma \gg \Gamma_{cr}$ and $\left|2\Omega_j - 2\Omega_i\right| \gg 2\delta_j + 2\delta_i$ are met, where $\delta_i \approx (\Gamma + \Gamma_{fb,i})/2$ is the linewidth of $S_{x_i}(\omega)$. Also, the unwanted force $\hat{z}\tilde{\rho}_{j3}$ does not affect parametric feedback cooling if the conditions $\Gamma \gg \Gamma_{cr}$ and $\left|2\Omega_j - \Omega_3\right| \gg 2\delta_j + \delta_3$ are met.

Let us now examine the effect of $\hat{x}_i\tilde{g}_{n,ji}$, which reads $-\hat{x}_i a_{ji}\eta_j(\dot{x}_j n_j \oplus x_j \dot{n}_j)x_i$. Unlike $\hat{x}_i\tilde{g}_{ji}$ (for $i \neq j$), $\hat{x}_i\tilde{g}_{n,ji}$ has a spectral density which always overlaps with the susceptibility of $x_i$, because $S_{n_j}(\omega)$ is a constant function of $\omega$. However, given the small linewidth of the susceptibility of $x_i$, the variance of $n_i$ as seen by the particle is much smaller than the variance of $x_i$ (unless the mechanical mode is cooled to near its ground state). Therefore, we can consider $\hat{x}_i\tilde{g}_{n,ji}$ as fluctuations, and use the quantum noise approach discussed in Section 3 to derive the rate $\tilde{\Gamma}_{g,i}$ associated with $\hat{x}_i\sum_j \tilde{g}_{n,ji}$ in the same way as the rate $\Gamma_{g,i}$ was derived. If we write $\eta_j$ as $M\Gamma_{fb,j}/\overline{x_j^2}$, and also approximate $S_{\dot{x}_j n_j \oplus x_j \dot{n}_j}(\omega)$ (viz., the spectral density of $\dot{x}_j n_j \oplus x_j \dot{n}_j$) by $\omega^2 \overline{x_j^2} S_{n_j}$, the rate $\tilde{\Gamma}_{g,i}$ is found to be equal to $\sum_j a_{ji}^2 \Gamma_{fb,j}^2 S_{n_j}/(4\overline{x_j^2})$, where $\overline{x_j^2}$ has yet to be found. We note that the intrinsic damping rate $\Gamma$ must be kept well above $8\tilde{\Gamma}_{g,i}$ for the same reason that it must be kept well above $8\Gamma_{g,i}$. This condition means that $\Gamma_{fb,j}$ (for each *j*) must be kept well below a critical value $\Gamma_{fb,cr,j}$.

The force $\hat{x}_3\tilde{\rho}_{n,j3}$, which reads $\hat{x}_3 b_j \eta_j(\dot{x}_j n_j \oplus x_j \dot{n}_j)$, can be considered as fluctuations as well. Therefore, we can use the quantum noise approach discussed in Section 3 to derive the rate $\tilde{\Gamma}_{r,3}$ associated with $\hat{x}_3\sum_j \tilde{\rho}_{n,j3}$ in the same way as the rates $\Gamma_{r,i}$ were derived. If we write $\eta_j$ as $M\Gamma_{fb,j}/\overline{x_j^2}$, and also approximate $S_{\dot{x}_j n_j \oplus x_j \dot{n}_j}(\omega)$ by $\omega^2 \overline{x_j^2} S_{n_j}$, the rate $\tilde{\Gamma}_{r,3}$ is found to be equal to $M\Omega_3 \sum_j b_j^2 \Gamma_{fb,j}^2 S_{n_j}/(2\hbar\overline{x_j^2})$, where $\overline{x_j^2}$ has yet to be found.

The rates $\tilde{\Gamma}_{r,1}$ and $\tilde{\Gamma}_{r,2}$ are zero. It should be noted that, due to the recoil force fluctuations, the rates $\Gamma_{r,1}$ and $\Gamma_{r,2}$ derived in Section 3 were not zero.

Having derived the rates $\tilde{\Gamma}_{g,i}$ and $\tilde{\Gamma}_{r,i}$, we can now find the mean phonon number $\overline{m}_i$ via Eq. (3) [or Eq. (4)]. However, we note that the rates $\tilde{\Gamma}_{g,i}$ and $\tilde{\Gamma}_{r,i}$ themselves depend on the mean phonon number $\overline{m}_i$, because they depend on $\overline{x_i^2}$, which is equal to $\overline{x}_i^{\,2} + (2\overline{m}_i + 1)\hbar/(2M\Omega_i) \approx \overline{m}_i\hbar/(M\Omega_i)$. The rates $\tilde{\Gamma}_{g,i}$ and $\tilde{\Gamma}_{r,i}$ also depend on the mean phonon number $\overline{m}_j$ (for $j \neq i$), because they depend on $\overline{x_j^2} \approx \overline{m}_j\hbar/(M\Omega_j)$. Therefore, the mean phonon numbers must be calculated self-consistently.

One can improve the performance of the system by filtering the electric current $I_j$ (or the electric current $I_j\dot{I}_j$) in a way that the information about $x_j$ (or $x_j\dot{x}_j$) remains intact while $n_j$ (or $\dot{x}_j n_j \oplus x_j\dot{n}_j$) converts into fluctuations whose spectral density is localized within a small enough linewidth around $\omega = \pm\Omega_j$ (or $\omega = \pm 2\Omega_j$). If such filtering is employed, not only the rate $\tilde{\Gamma}_{r,3}$ becomes zero, the rate $\tilde{\Gamma}_{g,i}$ also decreases and becomes equal to $\Gamma_{fb,i}^2 S_{n_i} / (4\overline{x_i^2})$. As a result, the condition that $8\tilde{\Gamma}_{g,i}$ (for all *i*) must be kept well below $\Gamma$ becomes equivalent to the condition that $\Gamma_{fb,i}^3$ must be kept well below a critical value $\Gamma_{fb,cr,i}^3$ equal to $(\overline{m}_{th,i}\Gamma + \Gamma_{r,i})\Gamma\hbar/(2M\Omega_i S_{n_i})$. In the derivation of the critical value $\Gamma_{fb,cr,i}^3$, Eq. (4) has been used. Also, $\Gamma_{fb,i}$ has been assumed to be much larger than $\Gamma$ (but $\Gamma_{fb,i}$ is much smaller than $\Gamma_{fb,cr,i}$).

In short, we can say that in the parametric feedback cooling scheme, the conditions (i) $|\Omega_j - \Omega_i| \gg \delta_j + \delta_i$ (for $j \neq i$), (ii) $|2\Omega_i - \Omega_3| \gg 2\delta_i + \delta_3$, (iii) $\Gamma \gg \Gamma_{cr}$, and (iv) $\Gamma_{fb,i}^3 \ll \Gamma_{fb,cr,i}^3$ must be met. Assuming that $\Gamma_{fb,i} \gg \Gamma$, the mean phonon number $\overline{m}_i$ is simplified to $(\overline{m}_{th,i}\Gamma + \Gamma_{r,i})/\Gamma_{fb,i}$ for $i$=1,2, and to $(\overline{m}_{th,3}\Gamma + \Gamma_{r,3} + \tilde{\Gamma}_{r,3})/\Gamma_{fb,3}$ for *i*=3. If the filtering described above is employed, $\tilde{\Gamma}_{r,3}$ is zero, and $\Gamma_{fb,cr,i}^3$ (for each *i*) is equal to $(\overline{m}_{th,i}\Gamma + \Gamma_{r,i})\Gamma\hbar/(2M\Omega_i S_{n_i})$. To minimize $\overline{m}_i$, we have to choose the maximum possible value well below $\Gamma_{fb,cr,i}$ for $\Gamma_{fb,i}$ (note that $\Gamma_{fb,cr,i}$ depends on $\Gamma$).

### *4.2 Hybrid feedback cooling*

More recently, the Coulomb force has been used in feedback systems of levitated particles [34-36]. In such systems, one component of the particle motion is cooled by the Coulomb force while the other two components are cooled by the optical gradient force in the way described in the previous subsection. To be specific, let us assume that the *k*th component of the particle motion is to be cooled by the Coulomb force.

Let us first examine the *j*th component of the cooling force, where $j \neq k$. This component, which is of the form $-\eta_j x_j^2 \dot{x}_j$, is accompanied by a gradient force $\hat{x}_i \tilde{g}_{ji}$ (for $i \neq j$), another gradient force $\hat{x}_i \tilde{g}_{n,ji}$ (for all *i*), and a radiation pressure $\hat{x}_3 (\tilde{\rho}_{j3} + \tilde{\rho}_{n,j3})$, where the expressions of $\tilde{g}_{ji}$, $\tilde{g}_{n,ji}$, $\tilde{\rho}_{j3}$, and $\tilde{\rho}_{n,j3}$ were given in the previous subsection. We saw that, to prevent $\tilde{g}_{ji}$ (for $i \neq j$) and $\tilde{\rho}_{j3}$ from disabling feedback cooling, (i) the intrinsic damping rate $\Gamma$ must be kept well above the critical value $\Gamma_{cr}$ derived in Section 3, and (ii) the conditions $\left|2\Omega_j - 2\Omega_i\right| \gg 2\delta_j + 2\delta_i$ (for $i \neq j$) and $\left|2\Omega_j - \Omega_3\right| \gg 2\delta_j + \delta_3$ must be met. Also, according to the previous subsection, the effect of $\hat{x}_i \sum_{j \neq k} \tilde{g}_{n,ji}$ on the mean phonon number $\overline{m}_i$ (for each *i*) is determined by the rate $\tilde{\Gamma}_{g,i}$, which is equal to $\sum_{j \neq k} a_{ji}^2 \Gamma_{fb,j}^2 S_{n_j} / (4\overline{x_j^2})$, while the effect of $\hat{x}_3 \sum_{j \neq k} \tilde{\rho}_{n,j3}$ on the mean phonon number $\overline{m}_3$ is determined by the rate $\tilde{\Gamma}_{r,3}$, which is equal to $M\Omega_3 \sum_{j \neq k} b_j^2 \Gamma_{fb,j}^2 S_{n_j} / (2\hbar \overline{x_j^2})$. We note that the intrinsic damping rate $\Gamma$ must be kept well above $8\tilde{\Gamma}_{g,i}$ (for all *i*).

According to the previous subsection, if we filter the electric current $I_j$ in a way that the spectral density of the resulting current is localized around $\omega = \pm\Omega_j$, (i) the rate $\tilde{\Gamma}_{g,k}$ becomes zero, (ii) the rate $\tilde{\Gamma}_{g,i}$ (for $i \neq k$) is reduced to $\Gamma_{fb,i}^2 S_{n_i} / (4\overline{x_i^2})$, and (iii) the rate $\tilde{\Gamma}_{r,3}$ becomes zero. In such a case, the condition that $8\tilde{\Gamma}_{g,i}$ (for all *i*) must be kept well below $\Gamma$ becomes equivalent to the condition that $\Gamma_{fb,i}^3$ (for $i \neq k$) must be kept well below the critical value $\Gamma_{fb,cr,i}^3 \approx (\overline{m}_{th,i}\Gamma + \Gamma_{r,i})\Gamma\hbar / (2M\Omega_i S_{n_i})$. Also, to minimize $\overline{m}_i \approx (\overline{m}_{th,i}\Gamma + \Gamma_{r,i}) / \Gamma_{fb,i}$ (for $i \neq k$), we have to choose the maximum possible value well below $\Gamma_{fb,cr,i}$ for $\Gamma_{fb,i}$.

Let us now examine the *k*th component of the cooling force. This component is supposed to be the Coulomb force, and therefore, requires that the particle has a net charge, and is trapped between the plates of a capacitor whose plates are perpendicular to the *k*th coordinate axis. Also, the second time-derivative of $I_k$ has to be applied to the capacitor, because the voltage of the capacitor and the resultant Coulomb force are proportional to the time-integral of the electric current applied to it ($I_k$ denotes the electric current generated by the photodetector intended to measure $x_k$). An implicit assumption is that the EM fields generated by the capacitor are quasi-static. This is a valid assumption especially if $I_k$ is filtered in a way that the spectral density of the resulting current is localized around $\omega = \pm\Omega_k$, because the dimensions of the capacitor are much smaller than the wavelength $\lambda_k = 2\pi c / \Omega_k$ corresponding to the mechanical angular frequency $\Omega_k$ (*c* denotes the speed of light in free space).

The *k*th component of the cooling force (viz., the component that is a Coulomb force) can be written as $-M\Gamma_{fb,k}\dot{x}_k$. This component is only accompanied by the force fluctuations $-\hat{x}_k M\Gamma_{fb,k}\dot{n}_k$. Therefore, the *k*th component of the cooling force is not accompanied by any force fluctuations that can contribute to $\tilde{\Gamma}_{g,i}$ (for any *i*). In other words, there is no critical value $\Gamma_{fb,cr,k}$ for $\Gamma_{fb,k}$ (viz., $\Gamma_{fb,cr,k}$ is infinite). Also, the *k*th component of the cooling force is not accompanied by any force fluctuations that can contribute to $\tilde{\Gamma}_{r,3}$. However, the Coulomb force fluctuations $-\hat{x}_k M\Gamma_{fb,k}\dot{n}_k$ lead to a non-zero $\tilde{\Gamma}_{r,k}$, which is equal to $M\Omega_k\Gamma_{fb,k}^2 S_{n_k} / (2\hbar)$. Although $\tilde{\Gamma}_{r,k}$ comes from the Coulomb force fluctuations (neither radiation pressure fluctuations nor recoil force fluctuations), we have used the subscript '*r*' for $\tilde{\Gamma}_{r,k}$ in order that we do not rewrite Eqs. (3) and (4) for the mean phonon number $\overline{m}_k$.

It should be noted that the Coulomb force has been assumed to be parallel to the *k*th coordinate axis. This is a valid assumption because the particle is far from the edges of the plates. However, even if the *k*th component of the cooling force was also accompanied by a Coulomb force $\hat{x}_i\xi_{ki}(\dot{x}_k + \dot{n}_k)$ along other coordinate axes (viz., for $i \neq k$), where $\xi_{ki}$ are certain coefficients, then (i) the force $\hat{x}_i\xi_{ki}\dot{x}_k$ would not disable feedback cooling because the conditions $\Gamma \gg \Gamma_{cr}$ and $\left|\Omega_k - \Omega_i\right| \gg \delta_k + \delta_i$ have already been met, and (ii) the force fluctuations $\hat{x}_i\xi_{ki}\dot{n}_k$ could be converted into fluctuations which would not increase $\overline{m}_i$ (by filtering the electric current $I_k$).

In short, we can say that in the hybrid feedback cooling scheme, where the *k*th component of the cooling force is the Coulomb force, the conditions (i) $\left|\Omega_j - \Omega_i\right| \gg \delta_j + \delta_i$ (for $j \neq i$), (ii) $\left|2\Omega_j - \Omega_3\right| \gg 2\delta_j + \delta_3$, (for $j \neq k$) (iii) $\Gamma \gg \Gamma_{cr}$, and (iv) $\Gamma_{fb,i}^3 \ll \Gamma_{fb,cr,i}^3$ (for $i \neq k$) must be met. There is no critical value $\Gamma_{fb,cr,k}$ for $\Gamma_{fb,k}$ (viz., $\Gamma_{fb,cr,k} = \infty$). Assuming that the electric currents are filtered, $\Gamma_{fb,cr,i}^3$ (for $i \neq k$) is equal to $(\overline{m}_{th,i}\Gamma + \Gamma_{r,i})\Gamma\hbar / (2M\Omega_i S_{n_i})$. Assuming that $\Gamma_{fb,i} \gg \Gamma$, the mean phonon number $\overline{m}_i$ is simplified to $(\overline{m}_{th,i}\Gamma + \Gamma_{r,i}) / \Gamma_{fb,i}$ for $i \neq k$, and to $(\overline{m}_{th,k}\Gamma + \Gamma_{r,k} + \tilde{\Gamma}_{r,k}) / \Gamma_{fb,k}$ for $i = k$. To minimize $\overline{m}_i$ for $i \neq k$, we have to choose the maximum possible value well below $\Gamma_{fb,cr,i}$ for $\Gamma_{fb,i}$. However, to minimize $\overline{m}_k$, we have to choose the optimum value $\Gamma_{fb,opt,k} = \sqrt{(\overline{m}_{th,k}\Gamma + \Gamma_{r,k})2\hbar / (M\Omega_k S_{n_k})}$ for $\Gamma_{fb,k}$.

The critical value $\Gamma_{fb,cr,i}$ (for $i \neq k$) as well as the optimum value $\Gamma_{fb,opt,k}$ are both manifestations of the fact that, due to the measurement noise, we cannot choose arbitrarily large values for the feedback cooling rates. While the existence of optimum feedback cooling rates in feedback systems (viz., feedback systems of non-levitated objects [21] as well as feedback systems of levitated particles [34]) has been underscored in other papers, the existence of critical feedback cooling rates, which is peculiar to feedback systems of levitated particles, has not so far been investigated.

We saw in Section 3 that the critical intrinsic damping rate $\Gamma_{cr}$ is insensitive to the mean power $\overline{P}_L$ of the Gaussian beam trapping the particle (and illuminating the particle for photodetection), and is also insensitive to the radius *R* of the particle, but is an increasing function of the numerical aperture *NA* of the lens employed to generate the Gaussian beam. Let us now see how the minimum mean phonon number $\overline{m}_{\min,k} = 2\sqrt{(\overline{m}_{th,k}\Gamma + \Gamma_{r,k})M\Omega_k S_{n_k} / (2\hbar)}$ (viz., $\overline{m}_k$ evaluated at $\Gamma_{fb,k} = \Gamma_{fb,opt,k}$ ) varies with $\overline{P}_L$ , *R*, and *NA*. The minimum mean phonon number $\overline{m}_{\min,k}$ is a decreasing function of $\overline{P}_L$ , because (i) $\overline{m}_{th,k}M\Omega_k S_{n_k} \propto TS_{n_k}$ is a decreasing function of $\overline{P}_L$ (although *T* is an increasing function of $\overline{P}_L$ ), and (ii) $\Gamma_{r,k}M\Omega_k S_{n_k} \propto \overline{P}_L S_{n_k}$ is insensitive to $\overline{P}_L$ (see Appendices C and D for the expressions of *T* and $S_{n_k}$ , respectively). The minimum mean phonon number $\overline{m}_{\min,k}$ is also a decreasing function of *R*, because (i) $\overline{m}_{th,k}M\Omega_k S_{n_k} \propto TMS_{n_k}$ is a decreasing function of *R* (although *M* is an increasing function of *R*), and (ii) $\Gamma_{r,k}M\Omega_k S_{n_k} \propto (B_k'^2 + C_k^2)S_{n_k}$ is insensitive to *R* (see Appendix B for the expressions of $B_k'$ and $C_k$ ). The minimum mean phonon number $\overline{m}_{\min,k}$ is a decreasing function of *NA*, because (i) $\overline{m}_{th,k}M\Omega_k S_{n_k} \propto TS_{n_k}$ is a decreasing function of *NA*, and (ii) $\Gamma_{r,k}M\Omega_k S_{n_k} \propto (B_k'^2 + C_k^2)S_{n_k}$ is insensitive to or a decreasing function of *NA* (depending on *k*). However, if we choose $\Gamma$ to be equal to $\xi\Gamma_{cr}$ (where $\xi$ is a certain number less than unity), we have to take into account the dependence of $\Gamma_{cr}$ on *NA*. In such a case, the minimum mean phonon number $\overline{m}_{\min,k}$ is still a decreasing function of *NA* when *k*=3, but is an increasing function of *NA* when *k*=1 or *k*=2. When choosing $\overline{P}_L$ , *R*, and *NA*, it should be noted that the surface temperature of the particle must remain below the melting point.

## 5. Numerical examples

Let us assume that the wavelength and the mean power of the Gaussian beam trapping the particle (and illuminating it for photodetection) are $\lambda_0$ =1064 nm and $\overline{P}_L$ =100 mW. The numerical aperture of the lens employed to generate the Gaussian beam is *NA*=0.8. The Gaussian beam propagates parallel to the *z* axis, and its electric field is polarized parallel to the *x* axis. The particle is of fused silica with a mass density of 2.2 gr/cm$^3$ and relative permittivity of 2.1+j10$^{-5}$.

The calculated mechanical oscillation frequencies $\Omega_1$ (and $\Omega_2$ ) and $\Omega_3$ are equal to 2π×367 KHz and 2π×208 KHz, respectively. We saw in the previous section that parametric feedback cooling and hybrid feedback cooling both require that $\Omega_i$ and $\Omega_j$ (for $j \neq i$) are not exactly equal, whereas the calculated $\Omega_1$ and $\Omega_2$ are equal because the Gaussian beam is symmetrical. Fortunately, in practice $\Omega_1$ and $\Omega_2$ are not exactly equal, because (i) the lens (and the resulting Gaussian beam) is not exactly symmetrical, and (ii) a Gaussian beam is not an exact solution to Maxwell's equations in the first place.

For a particle of radius *R*=70 nm, the calculated critical intrinsic damping rate $\Gamma_{cr}$ and its corresponding critical ambient pressure $P_{am,cr}$ are equal to 2π×791 nHz and $7\times10^{-10}$ mbar, respectively. We saw in Sections 3 and 4 that the ambient pressure ( $P_{am}$ ) must be kept well above $P_{am,cr}$ . Let us assume that $P_{am}$ is equal to $10P_{am,cr}$ =$7\times10^{-9}$ mbar. It is noteworthy that the ambient pressure chosen in a very recent experiment (where $\lambda_0$ =1064 nm, $\overline{P}_L$ =130 mW, *NA*=0.85, and *R*=68 nm) is equal to $7.5\times10^{-9}$ mbar [36].

The calculated surface temperature $T_s$ and the calculated temperature *T* (which determines $\overline{m}_{th,i}$ ) are equal to 1467 K and 697 K, respectively (note that they have been calculated at $P_{am} = 10P_{am,cr}$ =$7\times10^{-9}$ mbar). If we employ a larger particle of radius 180 nm, the calculated $\Omega_1$ , $\Omega_2$ , $\Omega_3$ , and $\Gamma_{cr}$ remain unchanged, but the calculated $P_{am,cr}$ , $T_s$, and *T* are now equal to $2\times10^{-9}$ mbar, 1857 K, and 866 K, respectively (note that $T_s$ and *T* have been now calculated at $P_{am} = 10P_{am,cr}$=$2\times10^{-8}$ mbar). The values 1467 K and 1857 K for $T_s$ are smaller than the melting point of fused silica, which is equal to 1873 K.

In Appendix D, the spectral density of the measurement noise has been derived in terms of parameters I named 'effective distance' (*Z*) and 'effective area' ( $a_d$ ). Let us assume that (i) the effective areas of the photodetectors are equal to their maximum allowable values [viz., $\lambda_0 Z / (45\pi)$ for the photodetectors intended to measure $x_1$ and $x_2$, and $\lambda_0 Z / (5\pi)$ for the photodetector intended to measure $x_3$ ], (ii) the effective distance between the photodetectors and the particle is as small as $10\lambda_0$, and (iii) the electric current generated by the photodetector intended to measure $x_i$ (for each *i*) is filtered in a way that the spectral density of the resulting current is localized around $\omega = \pm\Omega_i$ .

To examine the parametric feedback cooling scheme, let us first consider the particle of radius 70 nm at the ambient pressure of $7\times10^{-9}$ mbar. The calculated critical values $\Gamma^3_{fb,cr,1}$ and $\Gamma^3_{fb,cr,3}$ are equal to $(2\pi\times0.02\ \text{Hz})^3$ and $(2\pi\times0.3\ \text{Hz})^3$, respectively. We saw that in the parametric feedback cooling scheme, $\Gamma^3_{fb,i}$ (for each *i*) must be kept well below $\Gamma^3_{fb,cr,i}$ . Let us assume that $\Gamma^3_{fb,i}$ (for each *i*) is equal to $0.1\Gamma^3_{fb,cr,i}$ . In such a case, the calculated mean phonon numbers $\overline{m}_1$ and $\overline{m}_3$ are equal to $3\times10^4$ and $5\times10^3$, respectively. For the larger particle of radius 180 nm at the ambient pressure of 2×10$^{-8}$ mbar, the calculated $\Gamma^3_{fb,cr,1}$ and $\Gamma^3_{fb,cr,3}$ are equal to (2π×0.06 Hz)$^3$ and (2π×0.8 Hz)$^3$, respectively, and the calculated $\overline{m}_1$ and $\overline{m}_3$ are equal to 2×10$^4$ and 2×10$^3$, respectively.

The calculated $\overline{m}_3$ is considerably smaller than $\overline{m}_1$ (and $\overline{m}_2$), because the Gaussian beam illuminating the particle for photodetection propagates parallel to the *z* axis, and therefore, the spectral density of the measurement noise for $x_3$ is smaller than the corresponding values for $x_1$ (and $x_2$). Also, it should be noted that the calculated $\overline{m}_1$ and $\overline{m}_2$ are almost equal, because the only difference between the *x* and *y* components of the particle motion in our calculations is that, given the polarization of the Gaussian beam, the spectral density of the *x* component of the recoil force fluctuations is equal to half the corresponding value for the *y* component, and therefore, the rate $\Gamma_{r,1}$ is equal to half the rate $\Gamma_{r,2}$.

Let us now examine the hybrid feedback cooling scheme, and focus our attention on the *k*th component of the particle motion, which is the component to be cooled by the Coulomb force. For the particle of radius 70 nm at the ambient pressure of 7×10$^{-9}$ mbar, the calculated optimum feedback cooling rate $\Gamma_{fb,opt,k}$ and the calculated minimum mean phonon number $\overline{m}_{\min,k}$ are equal to 2π×2 Hz and 3×10$^2$ if *k*=1, and equal to 2π×94 Hz and 12 if *k*=3. For the particle of radius 180 nm at the ambient pressure of 2×10$^{-8}$ mbar, the calculated $\Gamma_{fb,opt,k}$ and $\overline{m}_{\min,k}$ are equal to 2π×10 Hz and 88 if *k*=1, and equal to 2π×465 Hz and 3 if *k*=3. For the same reasons as those given in the previous paragraph, (i) the calculated $\overline{m}_{\min,3}$ is considerably smaller than $\overline{m}_{\min,1}$ and $\overline{m}_{\min,2}$, and (ii) the calculated $\overline{m}_{\min,1}$ and $\overline{m}_{\min,2}$ are almost equal.

Our results suggest that the parametric feedback cooling scheme is not able to cool any component of the particle motion to near its ground state. Also, our results suggest that the hybrid feedback cooling scheme is able to cool the *z* component of the particle motion to near its ground state if the *z* component is cooled by the Coulomb force (viz., if *k*=3), but is not able to cool the *x* (or *y*) component to near its ground state even if the *x* (or *y*) component is cooled by Coulomb force (viz., even if *k*=1 or *k*=2). This is in agreement with the experimental results reported in [36] (where *k*=3) and [34] (where $k \neq 3$).

## 6. Concluding remarks

The true meaning of the optical gradient force and its corresponding operator was discussed in Section 3 (and Appendices A and B). Also, the spectral densities of radiation pressure and the recoil force were derived. In particular, it was shown that the approach used to derive the spectral densities leads to the same expectation value for radiation pressure as the result coming from the classical electrodynamics (viz., from the Maxwell stress tensor). Moreover, it was discussed how the spectral densities of radiation pressure and the recoil force depend on whether the laser power fluctuations are mainly due to the inherent uncertainty in the emission of photons by the laser or due to fluctuations in the electric current applied to the laser; the former case is relevant to the light trapping the particle, and the latter case is relevant to the light cooling the particle motion in the feedback cooling schemes.

It was shown that due to the inherent uncertainty in the emission of photons by the laser trapping the particle, and the resulting quantum fluctuations in the trapping force (which is a gradient force linear in the position of the particle), the particle escapes from its trap if the intrinsic damping rate (or, equivalently, the ambient pressure) becomes comparable with or smaller than a critical value. It was discussed how the renormalization effect leading to the particle's escape differs from the usual renormalization effect in cavity optomechanics. Also, it was argued that the cavity cooling and feedback cooling schemes realized so far cannot counteract that particle's escape at ambient pressures comparable with or smaller than the critical value. It was derived how the critical ambient pressure depends on the parameters of the system, especially on the mean laser power used to trap the particle, the radius of the particle, and the numerical aperture of the lens employed to generate the trapping beam.

Two main feedback cooling schemes (viz., parametric feedback cooling and hybrid feedback cooling) were analyzed in Section 4. To this end, the measurement noise coming from the uncertainty in the emission of photons by the laser illuminating the particle for photodetection (which is usually the same as the laser trapping the particle) was derived in terms of the parameters of the system, and the subtleties of the derivation were highlighted. It was then discussed that the measurement noise has two manifestations; one is well-known, and is relevant to feedback cooling systems of levitated particles and non-levitated objects both, and the other has not so far been investigated in the literature, and is relevant only to feedback cooling systems of levitated particles. The latter manifestation is that the particle escapes from its trap if any of the feedback cooling rates (corresponding to the three components of the particle motion in the parametric feedback cooling scheme, and corresponding to two components of the particle motion in the hybrid feedback cooling scheme) become comparable with or larger than critical values. The critical feedback cooling rates were derived by using a novel self-consistent method.

By giving some numerical examples, it was demonstrated that the parametric feedback cooling scheme is not able to cool any component of the particle motion to near its ground state. Also, it was shown that the hybrid feedback cooling scheme is able to cool the $z$ component of the particle motion to near its ground state if the $z$ component is cooled by the Coulomb force, but is not able to cool the $x$ (or $y$) component to near its ground state even if the $x$ (or $y$) component is cooled by Coulomb force, where we have assumed that the beam illuminating the particle for photodetection (which is usually the same as the beam trapping the particle) propagates parallel to the $z$ axis. These conclusions are in full agreement with the experimental results reported in the literature.

Finally, it was derived how the minimum mean phonon number (which is achieved in the hybrid feedback cooling scheme) for each component of the particle motion depends on the parameters of the system, especially on the mean laser power used to illuminate the particle for photodetection (which is usually the same as the mean laser power used to trap the particle), the radius of the particle, and the numerical aperture of the lens employed to generate the beam illuminating the particle for photodetection (which is usually the same as the beam trapping the particle).

## Appendix A: Classical EM force

The electric field of the Gaussian beam trapping the particle (and illuminating it for photodetection) reads

$$\vec{E}_L = \hat{x}_1 \,\mathrm{Re}(e^{-i\omega_0 t} E_{inc}) = \hat{x}_1 \,\mathrm{Re}\Big[ e^{-i\omega_0 t} \frac{E_0}{\sqrt{1 + X_3^2 / z_0^2}} e^{-\frac{X_1^2+X_2^2}{w_0^2(1+X_3^2/z_0^2)} + ik_0 X_3 + i\frac{k_0(X_1^2+X_2^2)}{2X_3(1+z_0^2/X_3^2)} - i\arctan(X_3/z_0)} \Big] \quad \text{(A1)}$$

where $(X_1, X_2, X_3)$ is the position of the observation point with respect to the focal point of the lens employed to generate the Gaussian beam, $z_0 = 2/(k_0 NA^2)$ is the Rayleigh range, $w_0 = 2/(k_0 NA)$ is the minimum beam radius, *NA* denotes the numerical aperture of the lens, $k_0 = 2\pi/\lambda_0 = \omega_0/c$ is the wavenumber, and *c* denotes the speed of light in free space [48]. The power carried by the Gaussian beam can be written as $P_L = \pi w_0^2 E_0^2/(4\eta_0)$ , where $\eta_0$ is the impedance of free space.

The dipole approximation assumes that the EM fields radiated by the particle, whose size is small in comparison with $\lambda_0$ , are almost equal to the EM fields radiated by a point-like dipole in free space [40,53,54]. By applying the dipole approximation to the Maxwell stress tensor [55], the classical EM force exerted by the Gaussian beam on the particle is simplified to $\vec{F} = 0.5\,\mathrm{Re}(\alpha E_{inc} \nabla E_{inc}^*)$ , where $\alpha$ denotes the polarizability of the particle, $E_{inc}$ is given by Eq. (A1), and $E_{inc}$ and $\nabla E_{inc}^*$ are both evaluated at the position of the particle center [40,57]. The EM force can be rewritten as $\vec{F} = \vec{g} + \vec{\rho}$ , where $\vec{g} = 0.25\alpha_R \nabla(|E_{inc}|^2)$ and $\vec{\rho} = 0.5\alpha_I \,\mathrm{Im}[E_{inc}^* \nabla(E_{inc})]$ are the so-called gradient force and radiation pressure, respectively, and $\alpha_R$ and $\alpha_I$ denote the real and imaginary parts of $\alpha$ . Assuming that the particle is a sphere of radius *R* and relative permittivity $\varepsilon$ , its polarizability reads $\alpha = \alpha_0 / [1 - ik_0^3 \alpha_0 / (6\pi\varepsilon_0)]$ , where $\alpha_0$ denotes $4\pi\varepsilon_0 R^3 (\varepsilon - 1)/(\varepsilon + 2)$ [40,53,54]. If the particle has a low loss [viz., when $\mathrm{Im}(\varepsilon) \ll \mathrm{Re}(\varepsilon)$ ], $\alpha_R$ and $\alpha_I$ can be approximated by $4\pi\varepsilon_0 a R^3$ and $8\pi\varepsilon_0 a^2 k_0^3 R^6 / 3$ , respectively, where $a$ denotes $(\varepsilon_R - 1)/(\varepsilon_R + 2)$ , and $\varepsilon_R$ denotes $\mathrm{Re}(\varepsilon)$ . It is noteworthy that $\vec{g}$ can also be derived by applying the dipole approximation to the method of virtual work rather than the Maxwell stress tensor [57].

Since the particle is around the focal point of the lens (viz., $|\vec{r}| \ll \lambda_0$ ), the calculated $\vec{g}$ can be approximated by a spring force $-(K_1 x_1, K_2 x_2, K_3 x_3)$ , where $\vec{r} = (x_1, x_2, x_3)$ is the position of the particle center. The stiffness vector $(K_1, K_2, K_3)$ can be written as $(A_1, A_2, A_3) P_L$ , where $A_3$ and $A_1 = A_2$ are found to be $aNA^6 k_0^4 R^3 / (2c)$ and $aNA^4 k_0^4 R^3 / c$ , respectively. Since the Gaussian beam in Eq. (A1) is symmetrical, $A_1$ and $A_2$ are equal. However, in practice $A_1$ and $A_2$ are unequal, because (i) the lens (and the resulting Gaussian beam) is not exactly symmetrical, and (ii) a Gaussian beam [like the one in Eq. (A1)] is not an exact solution to Maxwell's equations in the first place. The fact that $A_1$ and $A_2$ are unequal is fortunate, because, as is discussed in Section 4, what I name 'unwanted force components' disable parametric feedback cooling and hybrid feedback cooling if the oscillation frequencies $\Omega_i = \sqrt{K_i / M}$ and $\Omega_j = \sqrt{K_j / M}$ (for any $j \neq i$) are equal.

The calculated *z* component of radiation pressure ($\rho_3$) around the focal point of the lens is almost insensitive to $\vec{r}$, and can be written as $BP_L$, where $B$ is found to be $4a^2(1-0.5NA^2)NA^2k_0^6R^6/(3c)$. The components $\rho_1$ and $\rho_2$ around the focal point of the lens can be written as $\kappa_1 x_1$ and $\kappa_2 x_2$, respectively, where $\kappa_1$ and $\kappa_2$ are *positive*. Since $\kappa_1$ and $\kappa_2$ are much smaller than $K_1$ and $K_2$, we can ignore $\rho_1$ and $\rho_2$.

## Appendix B: EM force fluctuations

Let us use the same notation as in the main text and Appendix A. As was discussed in the main text, the gradient force is exerted on the particle by the EM modes *included* in the Hamiltonian, whereas interaction between the *system* (consisting of the particle and the EM modes included in the Hamiltonian) and the EM modes *excluded* from the Hamiltonian leads to radiation pressure and the recoil force. This is why we can (approximately) infer from the classical expression $g_i = -A_i P_L x_i$ derived in Appendix A that the Hamiltonian corresponding to the gradient force component $g_i$ reads $A_i P_L x_i^2 / 2$, where $P_L$ and $x_i$ are now interpreted as operators. In contrast, we can never infer from the classical expression $\overline{\rho}_3 = B\overline{P}_L$ derived in Appendix A that the Hamiltonian corresponding to radiation pressure reads $-BP_L x_3$ ($\overline{O} = \mathrm{E}\left[O\right]$ denotes the expectation value of *O*). We cannot even say that the spectral density of $\rho_3$ is necessarily equal to $B^2 S_{P_L}(\omega)$ (see [42] for the definition of spectral density). Also, the classical analysis in Appendix A is completely silent about the recoil force ($\vec{\sigma}$), whose expectation value is zero. In this appendix, we will first discuss the most general Hamiltonian we can write for the gradient force. We will then rigorously derive the spectral densities of $\rho_3$ and $\sigma_i$ (for *i*=1,2,3) in the case where the EM field fluctuations are mainly due to inherent uncertainty in the emission of photons by the laser – e.g. for the EM field trapping the particle (and illuminating it for photodetection in feedback cooling schemes). The case where the EM field fluctuations are mainly due to fluctuations in the electric current applied to the laser (e.g. for the EM field cooling the particle motion in feedback cooling schemes) will also be discussed.

### *B.1. Gradient force operator*

The gradient force is exerted on the particle by the EM modes included in the Hamiltonian. Here, a general Hamiltonian and operator corresponding to the gradient force will be presented. The Hamiltonians used in the literature on cavity optomechanics of levitated particles [23-28] are in fact special cases of the Hamiltonian to be presented.

Let us assume that our system consists of a certain number of EM modes (with bosonic operators $a_m$) which linearly interacts with each other and with the particle, whose position operator is denoted by $\vec{r}$. The Hamiltonian 'corresponding to the gradient force' can be written as

$$H = -0.25\alpha_R \sum_m e_m^2 \vec{\psi}_m^*(\vec{r}) \cdot \vec{\psi}_m(\vec{r}) a_m^\dagger a_m - 0.25\alpha_R \sum_m \sum_{n \neq m} e_m e_n \vec{\psi}_m^*(\vec{r}) \cdot \vec{\psi}_n(\vec{r}) a_m^\dagger a_n \quad , \tag{B1}$$

where the particle has been assumed to be lossless. The coefficient $\alpha_R$, which was defined in Appendix A, reads $4\pi\varepsilon_0 R^3(\varepsilon_R - 1)/(\varepsilon_R + 2)$ in terms of the radius ($R$) and relative permittivity ($\varepsilon_R$) of the particle. The function $\vec{\psi}_m$ has no physical dimension, and denotes the spatial electric field profile of the *m*th EM mode. The field profile $\vec{\psi}_m$ is normalized in the sense that the maximum of $\vec{\psi}_m^* \cdot \vec{\psi}_m$ is unity. The function $\vec{\psi}_m(\vec{r})$ merely reflects the fact that the electric field seen by the particle depends on the position of the particle.

If the *m*th EM mode is resonant (viz., a standing wave), then (i) the coefficient $e_m$ reads $\sqrt{2\hbar\omega_m/(\varepsilon_0 V_m)}$, where $V_m$ and $\omega_m$ denote the mode volume and the resonance frequency of the EM mode (in the absence of the particle), respectively, (ii) the operator $a_m$ obeys the equal-time commutation relation $[a_m(t), a_m^\dagger(t)] = 1$, and (iii) the operator $a_m^\dagger a_m$, which has no physical dimension, represents the number of photons in the EM mode.

If the *m*th EM mode is non-resonant (viz., a travelling wave), then (i) the coefficient $e_m$ reads $\sqrt{2\hbar\omega_L/(\varepsilon_0 c A_m)}$, where $\omega_L$ and $A_m$ denote the frequency of the driving laser and the mode area of the EM mode, respectively (e.g. $A_m$ is equal to $\pi w_0^2/2$ for the Gaussian beam introduced in Appendix A), (ii) the operator $a_m$ obeys the commutation relation $[a_m(t), a_m^\dagger(t')] = \delta(t - t')$, and (iii) the operator $a_m^\dagger a_m$, whose physical dimension is frequency, represents the rate of photons carried by the EM mode at a surface perpendicular to its propagation direction. It should be noted that the use of an operator like $s$ with the commutation relation $[s(t_1), s^\dagger(t_2)] = \delta(t_1 - t_2)$, which holds whether $t_1$ and $t_2$ are equal or not, means that the evolution of $s$ is known a priori. In other words, it means that the statistics of the non-resonant EM mode is determined by the laser, and is not affected by other EM modes or the particle motion. Strictly speaking, the operator $s$ with the commutation relation $[s(t_1), s^\dagger(t_2)] = \delta(t_1 - t_2)$ has to be replaced by $(1/\sqrt{2\pi})e^{-i\omega_L t}\int_{-\infty}^{\infty} b_\omega(t) e^{-i\omega t} d\omega$ in Eq. (B1), where the bosonic operators $b_{\omega_1}$ and $b_{\omega_2}$ (for any $\omega_1$ and $\omega_2$) obey the *equal-time* commutation relation $[b_{\omega_1}(t), b_{\omega_2}^\dagger(t)] = \delta(\omega_1 - \omega_2)$. The operator $s$ with the commutation relation $[s(t_1), s^\dagger(t_2)] = \delta(t_1 - t_2)$ is in fact equal to $(1/\sqrt{2\pi})e^{-i\omega_L t}\int_{-\infty}^{\infty} b_\omega(0)e^{-i\omega t}d\omega$ in terms of the operators $b_\omega$, where $t = 0$ is the time at which we know the state of the system. However, for problems such as those discussed in this paper, we can use operators like $s$ with the commutation relation $[s(t_1), s^\dagger(t_2)] = \delta(t_1 - t_2)$ for non-resonant EM modes in Eq. (B1), because in such problems we ignore the effect of the particle motion on the evolution of the non-resonant EM modes included in the system.

The gradient force operator is equal to the derivative of Eq. (B1) with respect to $\vec{r}$ with a minus sign, and can be written as

$$\begin{aligned} \vec{g} &= 0.25\alpha_R \sum_m e_m^2 \partial[\vec{\psi}_m^*(\vec{r}) \cdot \vec{\psi}_m(\vec{r})] / \partial\vec{r} a_m^\dagger a_m \\ &+0.25\alpha_R \sum_m \sum_{n \neq m} e_m e_n \partial[\vec{\psi}_m^*(\vec{r}) \cdot \vec{\psi}_n(\vec{r})] / \partial\vec{r} a_m^\dagger a_n \end{aligned} \qquad \text{(B2)}$$

It is worth noting that the evolution of the resonant EM modes of the system is interrelated with the evolution of $\vec{r}$, but, according to the definitions of 'Hamiltonian' and 'force', the derivatives of the operators $a_m$ with respect to $\vec{r}$ do not appear in Eq. (B2).

### *B.2. Spectral densities of radiation pressure and recoil force: Part One*

Let us assume that the EM field fluctuations are mainly due to inherent uncertainty in the emission of photons by the laser. In such a case, we can assume that the number of photons emitted by the laser in any time interval of length $\tau$ has a Poisson distribution with the expectation value $\overline{P}_L\tau/(\hbar\omega_0)$, and that the emission times are independent of each other, where $\omega_0$ and $\overline{P}_L$ denote the angular frequency and mean power of the laser, respectively. As a result, the spectral density of the laser power ($P_L$) is the constant function $S_{P_L}(\omega) = \hbar\omega_0\overline{P}_L$.

The radiation pressure ($\rho_3$) comes from the *initial* linear momentum of the photons interacting with the particle. The photons interacting with the particle are either scattered or absorbed by the particle. Let us write $\rho_3$ as the sum of $\hbar k_z \sum_{m=1}^{N'} \delta(t - t'_m)$ and $\hbar k_z \sum_{m=1}^{N''} \delta(t - t''_m)$, where $\hbar k_z$ is the initial linear momentum of the photons interacting with the particle, the observable $N'$ (or $N''$) is the number of the photons scattered (or absorbed) by the particle in the time interval $(0, T \to \infty)$, and the observables $t'_1, t'_2, ..., t'_{N'}$ (or $t''_1, t''_2, ..., t''_{N''}$) are the times at which the photons are scattered (or absorbed). Since the particle is around the focal point of the lens (viz., the expectation value of $|\vec{r}|$ is much smaller than $\lambda_0$ ), Eq. (A1) indicates that $k_z$ is equal to $k_0 - 1/z_0 = k_0(1-0.5NA^2)$ . Since the number of photons emitted by the laser in any time interval of length $\tau$ has a Poisson distribution with the expectation value $\overline{P}_L \tau/(\hbar\omega_0)$ , and since the emission times are independent of each other, we can say that: (i) $N'$ and $N''$ , which are independent of each other, have Poisson distributions with the expectation values $\overline{P}_s T/(\hbar\omega_0)$ and $\overline{P}_a T/(\hbar\omega_0)$ , respectively, where $\overline{P}_s = \mathrm{E}\left[P_s\right]$ and $\overline{P}_a = \mathrm{E}\left[P_a\right]$ are the mean optical powers scattered and absorbed by the particle, respectively, (ii) for a given $N'$ , the observables $t'_m$ and $t'_n$ (for $n \neq m$ ) are independent of each other, (iii) for a given $N''$ , the observables $t''_m$ and $t''_n$ (for $n \neq m$ ) are independent of each other, (iv) for a given $N'$ and $N''$ , the observables $t'_m$ and $t''_n$ are independent of each other, and (v) for a given $N'$ and $N''$ , the observables $t'_m$ and $t''_m$ each have a uniform distribution over the interval $(0,T)$ . Therefore, the expectation value of $\rho_3(t)\rho_3(t+\tau)$ can be written as

$$
\begin{aligned}
&\mathrm{E}\left[\rho_3(t)\rho_3(t+\tau)\right] = \mathrm{E}\left[\mathrm{E}\left[\rho_3(t)\rho_3(t+\tau)\,|\,N',N''\right]\right] = \\
&\hbar^2 k_z^2 \cdot \mathrm{E}\left[N' \int_0^T \delta(t-t')\delta(t-t'+\tau)\frac{dt'}{T}\right] + \\
&\hbar^2 k_z^2 \cdot \mathrm{E}\left[(N'^2 - N') \int_0^T \int_0^T \delta(t-t')\delta(t-\xi'+\tau)\frac{dt'd\xi'}{T^2}\right] + \\
&\hbar^2 k_z^2 \cdot \mathrm{E}\left[N'' \int_0^T \delta(t-t'')\delta(t-t''+\tau)\frac{dt''}{T}\right] + \\
&\hbar^2 k_z^2 \cdot \mathrm{E}\left[(N''^2 - N'') \int_0^T \int_0^T \delta(t-t'')\delta(t-\xi''+\tau)\frac{dt''d\xi''}{T^2}\right] + \\
&\hbar^2 k_z^2 \cdot \mathrm{E}\left[N'N'' \int_0^T \int_0^T \delta(t-t')\delta(t-t''+\tau)\frac{dt'dt''}{T^2}\right] + \\
&\hbar^2 k_z^2 \cdot \mathrm{E}\left[N''N' \int_0^T \int_0^T \delta(t-t'')\delta(t-t'+\tau)\frac{dt''dt'}{T^2}\right] = \\
&\frac{\hbar k_z^2 \overline{P}_s}{\omega_0}\delta(\tau) + \frac{k_z^2 \overline{P}_s^{\,2}}{\omega_0^2} + \frac{\hbar k_z^2 \overline{P}_a}{\omega_0}\delta(\tau) + \frac{k_z^2 \overline{P}_a^{\,2}}{\omega_0^2} + \frac{k_z^2 \overline{P}_s \overline{P}_a}{\omega_0^2} + \frac{k_z^2 \overline{P}_a \overline{P}_s}{\omega_0^2} = \\
&\frac{\hbar k_z^2 (\overline{P}_s + \overline{P}_a)}{\omega_0}\delta(\tau) + \frac{k_z^2 (\overline{P}_s + \overline{P}_a)^2}{\omega_0^2}
\end{aligned}
\tag{B3}
$$

where $\mathrm{E}\left[\rho_3(t)\rho_3(t+\tau)\,|\,N',N''\right]$ denotes the conditional expectation value of $\rho_3(t)\rho_3(t+\tau)$ given $N'$ and $N''$. The square root of the second term on the right hand side of Eq. (B3) is in fact the expectation value of $\rho_3$. Moreover, the Fourier transform of the first term on the right hand side of Eq. (B3) with respect to $\tau$ is the spectral density of $\rho_3$.

To derive $\overline{P}_s$ and $\overline{P}_a$, which appeared in Eq. (B3), we note that under the dipole approximation [40,53,54], the electric field radiated by the particle ($\vec{E}_r$) is equal to the electric field radiated by a point-like dipole with the electric dipole moment $\hat{x}_1\,\mathrm{Re}[e^{-i\omega_0 t}\alpha E_{inc}(\vec{r})]$, where $\alpha$ denotes the polarizability of the particle, and was given in Appendix A, and $E_{inc}(\vec{r})$ is the function given by Eq. (A1) and evaluated at $\vec{r}=(x_1,x_2,x_3)$. Since the particle is around the focal point of the lens (viz., the expectation value of $|\vec{r}\,|$ is much smaller than $\lambda_0$), its dipole moment can be approximated by $\hat{x}_1\,\mathrm{Re}[e^{-i\omega_0 t}\alpha E_0 e^{ik_z x_3}]$, where $k_z$ reads $k_0-1/z_0$. For a given $\vec{r}$, $\mathrm{E}\left[\vec{E}_r\times\vec{H}_r\right]$ in the far-field is found to be $|\alpha|^2\,NA^2k_0^6\overline{P}_L\sin^2(\theta')\hat{r}'/(32\pi^3\varepsilon_0^2r'^2)$, where $\mathrm{E}\left[E_0^\dagger E_0\right]$ has been replaced by $4\eta_0\overline{P}_L/(\pi w_0^2)$, and $(r',\theta',\varphi')$ denotes the position of the observation point with respect to the particle center in the spherical coordinate system whose zenith direction is parallel to the *x* axis. Therefore, $\overline{P}_s$, which is equal to the flux of $\mathrm{E}\left[\vec{E}_r\times\vec{H}_r\right]$ over any closed surface enclosing the particle, reads $|\alpha|^2\,NA^2k_0^6\overline{P}_L/(12\pi^2\varepsilon_0^2)$. Also, $\overline{P}_a$ is equal to $(\alpha_I\omega_0/2)\mathrm{E}\left[E_0^\dagger E_0\right]-\overline{P}_s$, where $\alpha_I$ denotes the imaginary part of $\alpha$. Therefore, if the particle has a low loss [viz., when $\varepsilon_I\ll\varepsilon_R$], $\overline{P}_s$ and $\overline{P}_a$ can be approximated by $4a^2NA^2k_0^6R^6\overline{P}_L/3$ and $6\varepsilon_I NA^2k_0^3R^3\overline{P}_L/(\varepsilon_R+2)^2$, respectively, where $\varepsilon_R+i\varepsilon_I$ is the relative permittivity of the particle, and $a$ denotes $(\varepsilon_R-1)/(\varepsilon_R+2)$. In such a case, $\overline{P}_a$ is much smaller than $\overline{P}_s$, and can be ignored.

Interestingly, the square root of the second term on the right hand side of Eq. (B3) is in agreement with $\overline{\rho}_3=B\overline{P}_L=4a^2(1-0.5NA^2)NA^2k_0^6R^6\overline{P}_L/(3c)$, which was derived in Appendix A by applying the dipole approximation to the Maxwell stress tensor. Moreover, the spectral density of $\rho_3$ [viz., the Fourier transform of the first term on the right hand side of Eq. (B3) with respect to $\tau$], can be written as $B'^2\,S_{P_L}(\omega)$, where $S_{P_L}(\omega)=\hbar\omega_0\overline{P}_L$ is the spectral density of $P_L$, and $B'$ is equal to $2a(1-0.5NA^2)NAk_0^3R^3/(\sqrt{3}c)$. It should be noted that $B'$ is not equal to $B$. Also, it should be noted that the expectation value and spectral density of $\rho_3$ are insensitive of the position of the particle in that the phase $k_z x_3$ in the expression of the dipole moment does not appear in $\vec{E}_r\times\vec{H}_r$. However, the phase $k_z x_3$ will be important in deriving the spectral density of the measurement noise in Appendix D.

The spectral densities of the components of the recoil force ( $\vec{\sigma}$ ) can be found in [32], but the derivation to be presented here, which is along the lines of Eq. (B3), allows us to make some important points. The recoil force ( $\vec{\sigma}$ ) comes from the *final* linear momentum of the photons scattered by the particle. Let us write $\sigma_i(t)$ as $-\sum_{m=1}^{N'} \hbar k'_{m,i}\delta(t-t'_m)$, where the observable $N'$ is the number of photons scattered by the particle in the time interval $(0, T\to\infty)$, the observables $t'_1, t'_2, ..., t'_{N'}$ are the times at which the photons are scattered, and the observable $k'_{m,i}$ is the *i*th component of the *final* wave vector ( $\vec{k}'_m$ ) of the *m*th photon scattered by the particle. One can write $\vec{k}'_{m,1}$, $\vec{k}'_{m,2}$, and $\vec{k}'_{m,3}$ as $k_0\cos(\theta'_m)$, $k_0\sin(\theta'_m)\cos(\varphi'_m)$, and $k_0\sin(\theta'_m)\sin(\varphi'_m)$, respectively, where the observables $\theta'_m$ and $\varphi'_m$ are the zenith and azimuth angles of $\vec{k}'_m$ in a spherical coordinate system whose zenith direction is parallel to the *x* axis. The observable $k'_{m,i}$ is independent of $t'_n$ (for all *n*), and is independent of $k'_{n,i}$ (for $n\neq m$ ). The observable $N'$ has a Poisson distribution with the expectation value $\overline{P}_s T/(\hbar\omega_0)$ . Also, for a given $N'$, the observables $t'_m$ and $t'_n$ (for $n\neq m$ ) are independent of each other, and have uniform distributions over the interval $(0,T)$ . Therefore, the expectation value of $\sigma_i(t)\sigma_i(t+\tau)$ can be written as

$$
\begin{aligned}
&\mathrm{E}\left[\sigma_i(t)\sigma_i(t+\tau)\right]=\mathrm{E}\left[\mathrm{E}\left[\sigma_i(t)\sigma_i(t+\tau)\,|\,N'\right]\right]=\\
&\hbar^2\mathrm{E}\left[k'^2_{m,i}\right]\cdot\mathrm{E}\left[N'\int_0^T\delta(t-t')\delta(t-t'+\tau)\frac{dt'}{T}\right]+\\
&\hbar^2\mathrm{E}^2\left[k'_{m,i}\right]\cdot\mathrm{E}\left[(N'^2-N')\int_0^T\!\!\int_0^T\delta(t-t')\delta(t-\xi'+\tau)\frac{dt'd\xi'}{T^2}\right]=\\
&\frac{\hbar\mathrm{E}\left[k'^2_{m,i}\right]\overline{P}_s}{\omega_0}\delta(\tau)+\frac{\mathrm{E}^2\left[k'_{m,i}\right]\overline{P}_s^{\,2}}{\omega_0^2}
\end{aligned}
\qquad\text{(B4)}
$$

Since $\mathrm{E}\left[k'_{m,i}\right]$ is zero, the second term on the right side of Eq. (B4) is zero, and therefore, the expectation value of $\sigma_i$ is zero. To calculate the spectral density of $\sigma_i$ [viz., the Fourier transform of the first term on the right hand side of Eq. (B4) with respect to $\tau$ ], $\mathrm{E}\left[k'^2_{m,i}\right]$ must be found. The spatial variation $\sin^2(\theta')$ of $\mathrm{E}\left[\vec{E}_r\times\vec{H}_r\right]$, which was discussed below Eq. (B3), means that the joint probably density function of $\theta'_m$ and $\varphi'_m$ in the expression of $\vec{k}'_m$ reads $3\sin^3(\theta'_m)/(8\pi)$, and, as a result, $\mathrm{E}\left[k'^2_{m,i}\right]$ is equal to $k_0^2/5$ for *i*=1, and to $2k_0^2/5$ for $i$=2,3. Therefore, the spectral density of $\sigma_i$ reads $C_i^2\hbar\omega_0\overline{P}_L$, where $C_i$ is equal to $2aNAk_0^3R^3/(\sqrt{15}c)$ for $i$=1, and to $2\sqrt{2}aNAk_0^3R^3/(\sqrt{15}c)$ for $i$=2,3. It should be noted that even if $\mathrm{E}\left[k'_{m,i}\right]$ was not zero, the second term on the right hand side of Eq. (B4) would not contribute to the spectral density of $\sigma_i$, because that term would still be independent of $\tau$. Therefore, it might seem that the fact $\mathrm{E}\left[k'_{m,i}\right]=0$ is always unimportant in deriving the spectral density of $\sigma_i$, but below we will see otherwise.

### *B.2. Spectral densities of radiation pressure and recoil force: Part Two*

Let us now assume that the EM field fluctuations are mainly due to fluctuations in the electric current applied to the laser. This is the case for the EM field cooling the particle motion in the feedback cooling schemes discussed in Section 4. Let us denote the laser power used to cool a certain component of the particle motion by $P_c$, and denote the corresponding current applied to the laser by $I_c$ (it is noteworthy that $I_c-\overline{I}_c$, which is here considered as fluctuations, is in fact the sum of the signal needed for feedback cooling and the measurement noise). We can derive an equation similar to Eq. (B3) for $\mathrm{E}\left[\rho_c(t)\rho_c(t+\tau)\right]$, where $\hat{z}\rho_c$ denotes the radiation pressure associated with the EM field supposed to cool that component of the particle motion. The first term on the right hand side of the equation for $\mathrm{E}\left[\rho_c(t)\rho_c(t+\tau)\right]$ will be similar to the first term on the right hand side of Eq. (B3), but can be ignored because $\overline{P}_c$ is small (viz., much smaller than $\overline{P}_L$). Unlike the second term on the right hand side of Eq. (B3), the second term on the right hand side of the equation for $\mathrm{E}\left[\rho_c(t)\rho_c(t+\tau)\right]$ will not be independent of $\tau$. To derive the second term, let us ignore absorption and make three assumptions: (i) the proportionality constant between $R_{P_c}(\tau)$ and $R_{I_c}(\tau)$ is equal to the proportionality constant between $\overline{P}_c^{\,2}$ and $\overline{I}_c^{\,2}$, where $R_O(\tau)$ denotes $\mathrm{E}[O(t)O(t+\tau)]-\overline{O}^{\,2}$, (ii) $R_{P_c}(\tau)$ is real, and (iii) the joint probability density function of $t'_{c,m}$ and $t'_{c,n}$ (for a given $N'$, and for $n\neq m$) over the interval $(0,T)$ is equal to $\mathbb{N}\left(1/T\right)^2[\overline{P}_c^{\,2}+R_{P_c}(t'_{c,m}-t'_{c,n})]$, where $t'_{c,1},t'_{c,2},\ldots,t'_{c,N'}$ denote the times at which photons of the cooling EM field are scattered by the particle, and $\mathbb{N}=1/[\overline{P}_c^{\,2}+(1/T)\int_0^T R_{P_c}(\tau)d\tau]$ is a normalizing constant [it is noteworthy that the joint probability density function which led to the second term on the right hand side of Eq. (B3) was $\left(1/T\right)^2$ ]. Under these assumptions, we can say that the proportionality constant between $R_{\rho_c}(\tau)$ and $R_{P_c}(\tau)$ is equal to the proportionality constant between $\overline{\rho}_c^{\,2}$ and $\overline{P}_c^{\,2}$. Also, the proportionality constant reads $B^2=[4a^2(1-0.5NA^2)NA^2k_0^6R^6/(3c)]^2$. Therefore, unlike the radiation pressure operator $\rho_3$ for the trapping EM field, which cannot be written as $BP_L$ in terms of the operator $P_L$, the radiation pressure operator $\rho_c$ for the cooling EM field can be approximated by $BP_c$ in terms of the operator $P_c$.

We can derive an equation similar to Eq. (B4) for $\mathrm{E}\left[\sigma_{c,i}(t)\sigma_{c,i}(t+\tau)\right]$, where $\sigma_{c,i}$ (for $i$=1,2,3) denote the components of the recoil force associated with the cooling EM field. The first term on the right hand side of the equation for $\mathrm{E}\left[\sigma_{c,i}(t)\sigma_{c,i}(t+\tau)\right]$ will be similar to the first term on the right hand side of Eq. (B4), but can be ignored because $\overline{P}_c$ is small (viz., much smaller than $\overline{P}_L$). Unlike the second term on the right hand side of Eq. (B4), the second term on the right hand side of the equation for $\mathrm{E}\left[\sigma_{c,i}(t)\sigma_{c,i}(t+\tau)\right]$ has factors dependent on $\tau$, but still does not contribute to the spectral density of $\sigma_{c,i}$ because $\mathrm{E}\left[k'_{c,m,i}\right]$ is zero, where $k'_{c,m,i}$ denotes the $i$th component of the final wave vector of the $m$th photon scattered by the particle. Therefore, the fact $\mathrm{E}\left[k'_{c,m,i}\right]=0$ is very important in allowing us to ignore the spectral density of $\sigma_{c,i}$.

## Appendix C: Thermal damping force and thermal random force

Let us first derive the surface temperature of the particle ($T_s$). Since the particle is very small and is around the focal point of the lens (viz., $R$ and the expectation value of $|\vec{r}\,|$ are much smaller than $\lambda_0$), and is also of rather high thermal conductivity, its surface temperature ($T_s$) is almost uniform, and is the solution to $\overline{P}_a = P_{c.c.} + P_{r.c.}$, where $\overline{P}_a$ denotes the mean optical power absorbed by the particle, $P_{c.c.}$ denotes the rate of heat conduction by the gas, and $P_{r.c.}$ denotes thermal radiation. The mean optical power absorbed by the particle ($\overline{P}_a$), which was derived in Appendix B, is equal to $6\varepsilon_I NA^2 k_0^3 R^3 \overline{P}_L / (\varepsilon_R + 2)^2$. The rate of heat conduction by the gas ($P_{c.c.}$) is equal to $\dfrac{(\gamma_a+1)P_{am}\overline{v}_{im}}{8(\gamma_a-1)T_{am}}\alpha_{acc}a_P(T_s-T_{am})$, where $\alpha_{acc}$ and $a_P = 4\pi R^2$ are the thermal accommodation coefficient and surface area of the particle, respectively, $\gamma_a$, $P_{am}$, and $T_{am}$ denote the heat capacity ratio, ambient pressure, and ambient temperature of the gas, respectively, and $\overline{v}_{im} = \sqrt{8k_B T_{am}/(\pi m)}$ is the mean velocity of the gas molecules impinging on the particle ($m$ denotes the mass of the gas molecules) [58,59]. The rate of thermal radiation ($P_{r.c.}$) is equal to $\sigma_0 a_P (T_s^4 - T_{am}^4)$, where $\sigma_0$ denotes the Stefan-Boltzmann constant, and the emissivity of the particle has been assumed to be unity.

The gas molecules surrounding the particle exert a damping force $-M\Gamma\vec{v}$ on the particle, where $M$ and $\vec{v}$ denote the mass and velocity of the particle, respectively. The intrinsic damping rate ($\Gamma$) is not given by Stokes' law, and is derived by using the kinetic theory of gases. According to Epstein's seminal paper [41], the intrinsic damping rate ($\Gamma$) for the motion of a spherical particle of radius *R* in a rarefied gas can be written as the sum of $\Gamma_{im} = \rho_a \overline{v}_{im} a_P / (3M)$, which is the contribution of the gas molecules impinging on the particle, and $\Gamma_{em} = \pi \rho_a \overline{v}_{em} a_P / (24M)$, which is the contribution of the gas molecules emerging from the particle, where $\rho_a = m_a P_{am} / (k_B T_{am})$ is the gas density, and $\overline{v}_{em} = \sqrt{8 k_B T_{em} / (\pi m)}$ is the mean velocity of the gas molecules emerging from the particle in terms of their temperature ($T_{em}$). Epstein's formula is applicable whenever the temperature $T_{em}$ is definable. The temperature $T_{em}$ can be written as $T_{am} + \alpha_{acc}(T_s - T_{am})$ [58,59].

Epstein's formula shows that the intrinsic damping rate ($\Gamma$) is almost insensitive to the mean laser power used to trap the particle ($\overline{P}_L$), and is almost proportional to the ambient pressure ($P_{am}$) and the inverse of the particle's radius ($1/R$). I say 'almost' because the temperature of the emerging molecules ($T_{em}$) itself depends on the surface temperature of the particle ($T_s$), and therefore depends on $\overline{P}_L$, $P_{am}$, and *R*.

The gas molecules also exert a random force $\vec{f}$ on the particle. This force is random even in the classical limit and even if the position and velocity of the particle are known. Also, this force is almost insensitive to the position and velocity of the particle. When the temperature of the emerging molecules ($T_{em}$) is equal to the ambient temperature ($T_{am}$), the spectral densities of the components of $\vec{f}$, which are equal to each other, can be derived by using the Caldeira-Leggett model, and written as a function $G(\omega; T_{am}, \Gamma)$ of $T_{am}$ and $\Gamma$ [60]. The function $G(\omega; T_{am}, \Gamma)$ can be rewritten as the sum of $G(\omega; T_{am}, \Gamma_{im})$ and $G(\omega; T_{am}, \Gamma_{em})$, where $\Gamma = \Gamma_{im} + \Gamma_{em}$, and the expressions of $\Gamma_{im}$ and $\Gamma_{em}$ were given above.

When the temperature of the emerging molecules ($T_{em}$) is *not* equal to the ambient temperature ($T_{am}$), the spectral density of the components of $\vec{f}$ can be written as the sum of $G(\omega; T_{am}, \Gamma_{im})$ and $G(\omega; T_{em}, \Gamma_{em})$, and simplified to $G(\omega; T, \Gamma)$, where *T* is defined as $(\Gamma_{im} T_{am} + \Gamma_{em} T_{em}) / \Gamma$ [44,45]. The temperature *T* is what appears in the expressions of $\overline{m}_{th,i}$ and $\overline{m}_i$ (for all *i*) in the main text.

**Appendix D: Measurement noise**

As is usually the case in experiments [31,32,34-36], let us assume that the beam trapping the particle illuminates it for photodetection as well. Let us use the same notations as in Appendix A and Appendix B. The measurement of the position of the particle is carried out by the measurement of the EM field intensity [31-36]. The electric field is the sum of the incident electric field ( $\vec{E}_L$ ) given by Eq. (A1) and the electric field radiated by the particle ( $\vec{E}_r$ ). Under the dipole approximation [40,53,54], $\vec{E}_r$ is equal to the electric field radiated by a point-like dipole with the electric dipole moment $\hat{x}_1 \mathrm{Re}[e^{-i\omega_0 t}\alpha E_{inc}(\vec{r})]$, where $\alpha$ denotes the polarizability of the particle, and $E_{inc}(\vec{r})$ is the function given by Eq. (A1) and evaluated at the particle center (viz., at $\vec{r}$ ). Since the particle is around the focal point of the lens (viz., the expectation value of $|\vec{r}|$ is much smaller than $\lambda_0$ ), its dipole moment can be approximated by $\hat{x}_1 \mathrm{Re}[e^{-i\omega_0 t}\alpha E_0 e^{ik_z x_3}]$, where $\vec{r}$ has been written as $(x_1, x_2, x_3)$, and $k_z$ reads $k_0 - 1/z_0$. It is noteworthy that the phase $k_z x_3$ in the expression of the dipole moment was not important in deriving the spectral densities of $\rho_3$ and $\sigma_3$, but it is now important in deriving $S_{n_3}$ (viz., the spectral density of the measurement noise corresponding to the measurement of $x_3$ ).

At an observation point $(X, Y, Z)$ far enough from the particle and close enough to the axis of the beam [viz., $Z > 10\lambda_0$ and $X^2, Y^2 < \lambda_0 Z / (20\pi)$ ], the sum of $\vec{E}_L$ and $\vec{E}_r$ can be written as

$$\begin{aligned}\vec{E}_d &= \hat{x}_1 \mathrm{Re}\Big\{ e^{-i\omega_0 t}(-i)E_0 \\ &\Big[ z_0 / Z + a k_0^3 R^3 X x_1 / Z^2 + a k_0^3 R^3 Y x_2 / Z^2 + a k_0^2 R^3 x_3 / (z_0 Z) \Big] \Big\}\end{aligned} \quad \text{(D1)}$$

where *a* and *R* denote $(\varepsilon_R - 1)/(\varepsilon_R + 2)$ and the radius of the particle, respectively, and $z_0$ denotes the Rayleigh range of the Gaussian beam illuminating the particle. It should be noted that $(X, Y, Z)$ and $\vec{r} = (x_1, x_2, x_3)$ have been defined with respect to the focal point of the lens employed to generate the Gaussian beam. Equation (D1) is in fact the electric field calculated within the dipole approximation, the far-field approximation, the paraxial approximation, and the assumption that the particle is around the focal point of the lens.

The electric field *operator* at $(X, Y, Z)$ is also given by Eq. (D1) if $\vec{r} = (x_1, x_2, x_3)$ and $E_0$ are interpreted as operators. The operator corresponding to the optical power carried by the beam can be written as $P_L = \pi w_0^2 E_0^\dagger E_0 / (4\eta_0)$ in terms of the operator $E_0$, where $w_0$ denotes the minimum beam radius of the beam, and $\eta_0$ denotes the impedance of free space.

The spectral density of the photocurrent $I_i$ generated by a small enough photodetector centered at $(X_i, Y_i, Z_i)$ can be written as the sum of $S_{M_i}(\omega)$ and $S_{N_i}(\omega)$, where $S_{N_i}(\omega)$, which is independent of $\omega$ over the detection bandwidth, is equal to $\zeta_i q^2 \mathrm{E}\left[E_{d_i}^{\dagger} E_{d_i}\right]$, and $S_{M_i}(\omega)$ is equal to the Fourier transform of

$$R_{M_i}(\tau) = \zeta_i^2 q^2 \mathrm{E}\left[E_{d_i}^{\dagger}(t) E_{d_i}^{\dagger}(t+\tau) E_{d_i}(t+\tau) E_{d_i}(t)\right] - \zeta_i^2 q^2 \mathrm{E}\left[E_{d_i}^{\dagger} E_{d_i}^{\dagger} E_{d_i} E_{d_i}\right] \tag{D2}$$

with respect to $\tau$ ( $\mathrm{E}\left[O\right]$ denotes the expectation value of *O*) [61]. The subscript '*i*' emphasizes that the photodetector is intended to measure $x_i$. The role of the second term on the right hand side of Eq. (D2) is to eliminate the $\tau$ -independent part of the first term. The coefficient *q* denotes the elementary charge, and the coefficient $\zeta_i$ reads $a_{d_i} / (2\eta_0 \hbar \omega_0)$ in terms of the area $a_{d_i}$ of the photodetector. The detection efficiency has been assumed to be unity, and the detection bandwidth has been assumed to be large in comparison with the mechanical oscillation frequency $\Omega_i$. Given that $S_{I_i}(\omega)$ is the sum of $S_{M_i}(\omega)$ and $S_{N_i}(\omega)$, we can interpret the fluctuations $I_i - \overline{I}_i$ as an incoherent sum of a signal $M_i$ with the spectral density $S_{M_i}(\omega)$ and a noise $N_i$ with the spectral density $S_{N_i}(\omega)$. Let us write $I_i - \overline{I}_i$ as $M_i \oplus N_i$.

Given Eq. (D1) and the fact that the expectation value of $|\vec{r}\,|$ is much smaller than $\lambda_0$, $S_{N_i}(\omega)$, which is independent of $\omega$, is found to be $\zeta_i q^2 \mathrm{E}\left[E_0^{\dagger} E_0\right] (z_0 / Z_i)^2$. Let us now derive $S_{M_i}(\omega)$ for each *i*.

The photodetector intended to measure $x_3$ is centered at $(X_3, Y_3, Z_3) = (0, 0, Z)$, and thus we can ignore any signature of $x_1$ and $x_2$ in $M_3$. Ignoring $(a k_0^2 R^3 / (z_0 Z))^4 \mathrm{E}\left[x_3(t) x_3^2(t+\tau) x_3(t) - x_3^4\right]$, we find that $S_{M_3}(\omega)$ can be written as $\theta_3 S_{x_3}(\omega)$, where the coefficient $\theta_3$ reads $(2\zeta_3 q a k_0^2 R^3 / Z^2)^2 \mathrm{E}\left[E_0^{\dagger 2} E_0^2\right]$. Therefore, $S_{n_3}$, defined in the main text, is found to be

$$S_{n_3} = S_{N_3} / \theta_3 = \frac{\pi \hbar \omega_0 w_0^2 z_0^2}{8 a^2 k_0^4 R^6 \overline{P}_L} \frac{Z^2}{a_{d_3}}. \tag{D3}$$

The expression of $S_{n_3}$ indicates that it is advantageous to increase $a_{d_3}$. However, in view of the paraxial approximation, the maximum allowable value of $a_{d_3}$ is around $\lambda_0 Z / (5\pi)$ at a given $Z$. If we assume that $a_{d_3}$ is equal to its maximum allowable value for any given $Z$, $S_{n_3}$ is proportional to $Z$. Therefore, it is advantageous to decrease $Z$. However, in view of the far-field approximation, $Z$ must be kept well above $\lambda_0$ (viz., $Z > 10\lambda_0$ ).

To measure $x_1$, the photocurrents $I_1 = \overline{I}_1 + M_1 \oplus N_1$ and $I_1' = \overline{I}_1' + M_1' \oplus N_1'$ generated by two photodetectors centered at $(X_1, Y_1, Z_1) = (X, 0, Z)$ and $(X_1', Y_1', Z_1') = (-X, 0, Z)$, respectively, are subtracted from each other [31]. Such a balanced detection allows us to ignore any signature of $x_3$ (and $x_2$) in $I_1 - I_1'$. It should be noted that $I_1 - I_1'$ cannot be written as $(\overline{I}_1 - \overline{I}_1') + (M_1 - M_1') \oplus (N_1 - N_1')$. In fact, $M_1$ and $-M_1'$ are added coherently (viz., they are perfectly correlated) while $N_1$ and $-N_1'$ are added incoherently (viz., they are uncorrelated). Therefore, let us write $I_1 - I_1'$ as $(\overline{I}_1 - \overline{I}_1') + (M_1 - M_1') \oplus [N_1 \oplus (-N_1')]$. The spectral density of $N_1 \oplus (-N_1')$ is equal to $2S_{N_1}(\omega)$. Ignoring $(ak_0^3 R^3 X / Z^2)^4 \mathrm{E}\left[ x_1(t) x_1^2(t+\tau) x_1(t) - x_1^4 \right]$, we find that the spectral density of $(M_1 - M_1')$ can be written as $4\theta_1 S_{x_1}(\omega)$, where $\theta_1$ reads $(2\zeta_1 q a k_0^3 R^3 z_0 X / Z^3)^2 \mathrm{E}\left[ E_0^{\dagger 2} E_0^2 \right]$. Therefore, $S_{n_1}$, defined in the main text, reads

$$S_{n_1} = 0.5\ S_{N_1} / \theta_1 = 0.5 \frac{\pi \hbar \omega_0 w_0^2}{8a^2 k_0^6 R^6 \overline{P}_L} \frac{Z^4}{X^2 a_{d_1}}. \qquad \text{(D4)}$$

The expression of $S_{n_1}$ indicates that it is advantageous to increase $X^2 a_{d_1}$. However, in view of the paraxial approximation, the maximum allowable value of $X^2 a_{d_1}$ is around $[\lambda_0 Z / (45\pi)]^2$ at a given $Z$. It happens when $X^2\ \&\ a_{d_1} \approx \lambda_0 Z / (45\pi)$. If we assume that $X^2\ \&\ a_{d_1} \approx \lambda_0 Z / (45\pi)$ for any given $Z$, $S_{n_1}$ is proportional to $Z^2$. Therefore, it is advantageous to decrease $Z$. However, in view of the far-field approximation, $Z$ must be kept well above $\lambda_0$ (viz., $Z > 10\lambda_0$).

To measure $x_2$, the photocurrents generated by two photodetectors centered at $(X_2, Y_2, Z_2) = (0, Y, Z)$ and $(X_2', Y_2', Z_2') = (0, -Y, Z)$ are subtracted from each other. The term $S_{n_2}$, defined in the main text, is derived in the same way as $S_{n_1}$ was derived. It reads

$$S_{n_2} = 0.5\ S_{N_2} / \theta_2 = 0.5 \frac{\pi \hbar \omega_0 w_0^2}{8a^2 k_0^6 R^6 \overline{P}_L} \frac{Z^4}{Y^2 a_{d_2}}. \qquad \text{(D5)}$$

In view of the paraxial approximation, the maximum allowable value of $Y^2 a_{d_2}$ is around $[\lambda_0 Z / (45\pi)]^2$ at a given $Z$. It happens when $Y^2\ \&\ a_{d_2} \approx \lambda_0 Z / (45\pi)$. If we assume that $Y^2\ \&\ a_{d_2} \approx \lambda_0 Z / (45\pi)$ for any given $Z$, $S_{n_2}$ is proportional to $Z^2$.

It is noteworthy that there was a subtle approximation involved in deriving $S_{M_i}(\omega)$ for each *i*. The correlation $\mathrm{E}\left[x_i(t)x_i(t+\tau)\right]$ has been approximated by its real part. In other words, $\mathrm{E}\left[E_{d_i}^{\dagger}(t)E_{d_i}^{\dagger}(t+\tau)E_{d_i}(t+\tau)E_{d_i}(t)\right]$ has been approximated by $\mathrm{E}\left[E_{d_i}^{\dagger}(t)E_{d_i}(t)E_{d_i}^{\dagger}(t+\tau)E_{d_i}(t+\tau)\right]$. The latter is in fact proportional to $\mathrm{E}\left[P_{d_i}(t)P_{d_i}(t+\tau)\right]$, where $P_{d_i}$ denotes the optical power received by the photodetector *i*. This approximation was necessary in deriving the relations $S_{M_i}(\omega)=\theta_i S_{x_i}(\omega)$.

Finally, it should be noted that in practice the photodetectors cannot be placed at a small distance of $10\lambda_0$ from the particle – rather, they are at a distance of $10^6\lambda_0 - 10^7\lambda_0$ from the particle. The effect of such a long distance on $S_{n_i}$ can be compensated to some extent by employing a collimating lens with a large enough numerical aperture to collect a large enough amount of light before we direct the light to the photodetectors. The focal point of the collimating lens must coincide with the focal point of the lens employed to generate the beam illuminating the particle. Also, each photodetector must be placed at the focal plane of a converging lens supposed to compensate the Fourier transforming effect of the collimating lens. In such a detection scheme, the results derived in this appendix all remain valid, but $Z$ no longer denotes the *actual* distance between the photodetectors and the focal point of the lens employed to generate the beam illuminating the particle. Also, $a_{d_i}$ no longer denotes the *actual* area of the photodetector intended to measure $x_i$. Rather, $Z$ is a parameter determined by the characteristics of the collimating lens, the converging lens, and other optical devices between them (e.g., beam splitters and mirrors), hence the name '*effective* distance' for $Z$ in the main text. The parameter $Z$ must still meet the condition $Z \gg \lambda_0$. Also, the maximum allowable values of the parameters $a_{d_3}$, $X^2 a_{d_1}$, and $Y^2 a_{d_2}$ are still around $\lambda_0 Z/(5\pi)$, $[\lambda_0 Z/(45\pi)]^2$, and $[\lambda_0 Z/(45\pi)]^2$, respectively, at a given $Z$. If we assume that $a_{d_3} \approx \lambda_0 Z/(5\pi)$ for any given $Z$, $S_{n_3}$ is still proportional to $Z$. If we assume that $X^2 \,\&\, Y^2 \,\&\, a_{d_1} \,\&\, a_{d_2} \approx \lambda_0 Z/(45\pi)$ for any given $Z$, $S_{n_1}$ and $S_{n_2}$ are still proportional to $Z^2$.

## Appendix E: Summary of important symbols

**Table 1. Definitions of important symbols used in the article**

| | |
|---|---|
| $\vec{r}=(x_1,x_2,x_3)$ | The position (or position operator) of the particle center with respect to the focal point of the lens |

| $\Gamma$ | The particle motion's intrinsic damping rate coming from the gas molecules surrounding the particle |
|---|---|
| $P_{am}$ and $T_{am}$ | The ambient pressure and the ambient temperature |
| $R$, $M$, and $\alpha_{acc}$ | The radius, the mass, and the accommodation coefficient of the particle |
| $T_s$ and $T$ | The surface temperature and the motional temperature of the particle (in the absence of EM fluctuations and also in the absence of feedback cooling) |
| $c$, $\omega_0$, $\lambda_0$, $\overline{P}_L$, and $NA$ | The speed of light in free space, the angular frequency, the wavelength, and the mean power of the Gaussian beam trapping the particle and illuminating it for photodetection, and the numerical aperture of the lens employed to generate the Gaussian beam |
| $\Omega_i$ | The mechanical oscillation angular frequency for the *i*th component of the particle motion |
| $g_i$, $\rho_i$, and $\sigma_i$ | The *i*th components of the gradient force operator, radiation pressure operator, and recoil force operator |
| $\varepsilon$, $\varepsilon_R$, $\varepsilon_I$, and $a = (\varepsilon_R - 1)/(\varepsilon_R + 2)$ | The relative permittivity of the particle, its real and imaginary parts, and a coefficient in terms of its real part |
| $\alpha$, $\alpha_R$, $\alpha_I$, and $\alpha_0$ | The polarizability of the particle, its real and imaginary parts, and the polarizability at zero frequency (but at the same permittivity) |
| $A_i$ | A coefficient that appears in the *i*th component of the classical gradient force and its quantum operator |
| $B$ | A coefficient that appears in the classical radiation pressure and also in the spectral density of the radiation pressure fluctuations coming from the measurement noise |
| $B'$ | A coefficient that appears in the spectral density of the radiation pressure fluctuations coming from the trapping EM field fluctuations |
| $C_i$ | A coefficient that appears in the spectral density of the *i*th component of the recoil force fluctuations coming from the trapping EM field fluctuations |
| $(X_i, Y_i, Z_i)$ , $a_{d,i}$ , and $n_i$ | The effective position of the photodetector intended to measure $x_i$ with respect to the focal point of the lens employed to generate the Gaussian beam illuminating the particle, the effective area of the photodetector, and the measurement noise |
| $\Gamma_{cr}$ and $P_{am,cr}$ | The critical intrinsic damping rate and the critical ambient pressure |
| $\Gamma_{fb,i}$ , $\Gamma_{fb,opt,i}$ , and $\Gamma_{fb,cr,i}$ | The feedback cooling rate, the optimum feedback cooling rate, and the critical feedback cooling rate, all for the *i*th component of the particle motion |
| $\overline{m}_i$ and $\overline{m}_{\min,i}$ | The mean phonon number and the minimum mean phonon number for the *i*th component of the particle motion |

## Acknowledgement

I am very grateful to Prof. Khashayar Mehrany, who authorized me to initiate the field of optomechanics in the Electrical Engineering Department at Sharif University of Technology.